\pdfoutput=1
\RequirePackage{fix-cm}

\documentclass[smallextended,a5paper]{svjour3}       

\smartqed  

\usepackage{graphicx}
\advance\paperwidth by-31mm
\advance\paperheight by-43.5mm
\usepackage{crop}
\papertype{generic article}

\usepackage{amsmath}
\usepackage{amssymb}
\usepackage{natbib}

\begin{document}

\title{The 64 Mpixel wide field imager for the Wendelstein 2m Telescope: Design and Calibration}

\titlerunning{Design and Calibration of the Wendelstein Wide Field Imager}        

 \author{Ralf\, Kosyra \and
          Claus\, G{\"o}ssl   \and
          Ulrich\, Hopp          \and
          Florian\, Lang-Bardl    \and
          Arno\, Riffeser      \and
          Ralf\, Bender        \and
          Stella\, Seitz
 }

 \institute{Ralf Kosyra \and Claus\, G{\"o}ssl \and Ulrich\, Hopp \and  Florian\, Lang-Bardl \and
 Arno\, Riffeser \and Ralf\, Bender \and Stella\, Seitz \at
              Universit{\"a}ts-Sternwarte M{\"u}nchen,
 	      Scheinerstra{\ss}e~1, D81679~M{\"u}nchen, Germany\\              
              \and 
              Ulrich\, Hopp \and Ralf Bender \and Stella\, Seitz 
              \at
              Max Planck Institut f{\"u}r Extraterrestrische Physik, Giessenbachstr., 
              D85748 Garching, Germany\\
              \\
              \email{kosyra@usm.uni-muenchen.de}
 }

\date{Received: 2014 May 27 / Accepted: 2014 August 10 \\To appear in Springer Experimental Astronomy}

\maketitle

\begin{abstract}
The Wendelstein Observatory of Ludwig Maximilians University of Munich 
has recently been upgraded with a modern 2m robotic telescope. One
Nasmyth port of the telescope has been equipped with a wide-field corrector 
which preserves the excellent image quality ($< 0.8"$ median seeing)  of the 
site \citep{2008SPIE.7016E..58H} over a field of view of 0.7 degrees diameter. 
The available field is imaged by an optical imager (WWFI, the Wendelstein Wide Field Imager)
built around a customized $2 \times 2$  mosaic of $4k \times 4k \ 15~\mu$m e2v CCDs from Spectral Instruments. 
This paper provides an overview of the design and the WWFI's performance.
We summarize the system mechanics (including a structural analysis), the electronics
(and its electromagnetic interference (EMI) protection) and the control software.
We discuss in detail detector system parameters, i.e. gain and readout noise, quantum efficiency as well as 
charge transfer efficiency (CTE) and persistent charges. First on sky tests yield
overall good predictability of system throughput based on lab measurements.
\end{abstract}

\section{Introduction}

\begin{figure*}[ht]
  \centering
  \includegraphics[width=\textwidth]{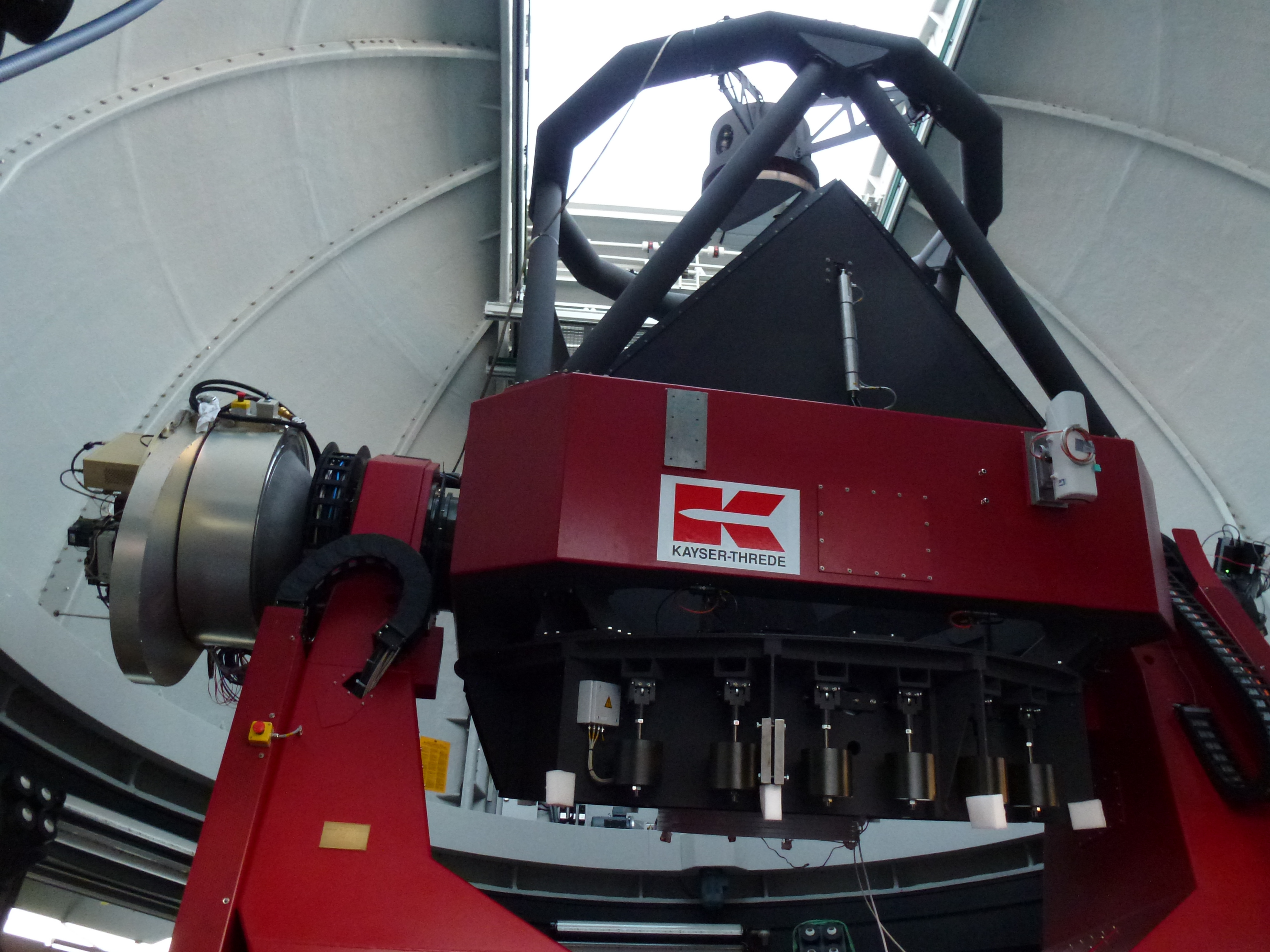}    
   \caption{WWFI (left) mounted at one Nasmyth port of the {\em Fraunhofer Telescope}.}
  \label{WWFI_telescope}
\end{figure*}

The Wendelstein Wide Field Imager (WWFI) was chosen as the scientific
first light instrument for the new { \em Fraunhofer Telescope} \footnote{{ \em Fraunhofer Telescope}
was built by Kayser-Threde GmbH, Munich and Astelco Systems GmbH, Martinsried} for two
reasons.
First, it should support the tedious alignment of  the very compact
optical system of the telescope, and second, it should provide early
science verification during telescope commissioning with a number of
projects we were already pursuing.
These projects were:
Difference imaging of Local Group galaxies to search for variables and
microlensing events
\citep[e.g.][]{2012AJ....143...89L,2013AJ....145..106K},
planet transit analyses
\citep[e.g.][]{2013MNRAS.435.3133K},
surface photometry of galaxies
\citep[e.g.][]{2012ApJS..198....2K}
and weak lensing mass estimates for galaxy clusters
\citep[e.g.][]{2013arXiv1310.6744G}.

 In Sect.~\ref{Components} we present an overview of the optical and
mechanical layout with the camera subcomponents, as well as the
electrical and software design of the WWFI.
In Sect.~\ref{LabComm} we describe all measurements
that have been performed in the laboratory to characterize the most
important parameters of the detector system:
Gains, the detectors' quantum efficiencies (QE), the readout
noises, the charge transfer efficiencies as well as the
characteristics of charge persistence of the CCDs are derived from
those data.
We also compare our results of combined lab efficiency measurements of
all optical elements and detectors with on sky commissioning
observations of globular cluster Messier 13 and three standard star
fields from the Landolt catalog
\citep{1973AJ.....78..959L,1983AJ.....88..439L,1992AJ....104..340L,2009AJ....137.4186L}.
In Sect.~\ref{sectComp} we compare our system to ESO OmegaCAM \citep{2006SPIE.6276E...9I} and
ESO WFI \citep{1999Msngr..95...15B} and conclude with a summary of our results in
Sect.~\ref{conc}.

\begin{figure*}
  \centering  
  \includegraphics[height=0.28\textheight]{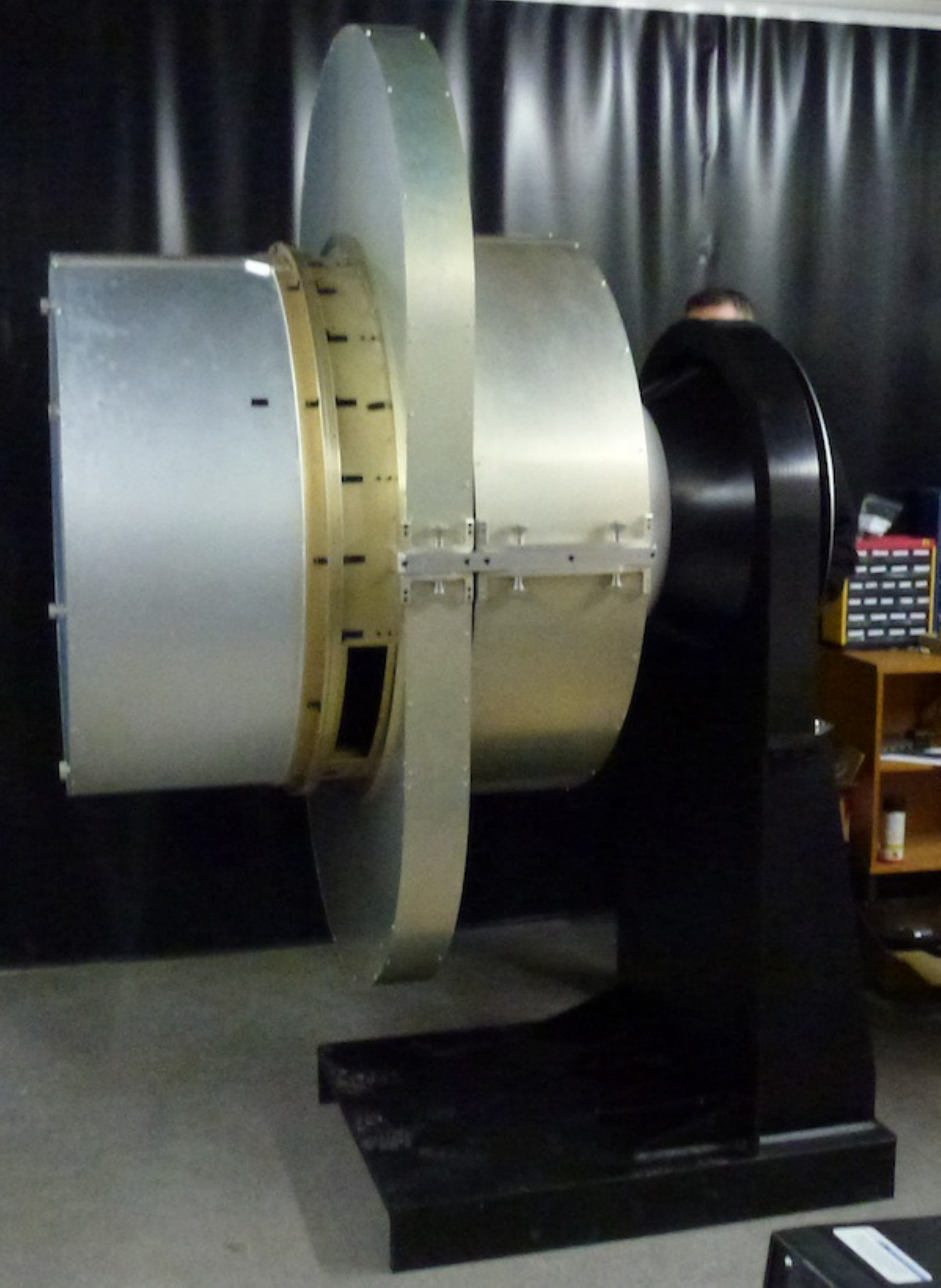} 
  \hspace*{2ex}
  \includegraphics[height=0.28\textheight]{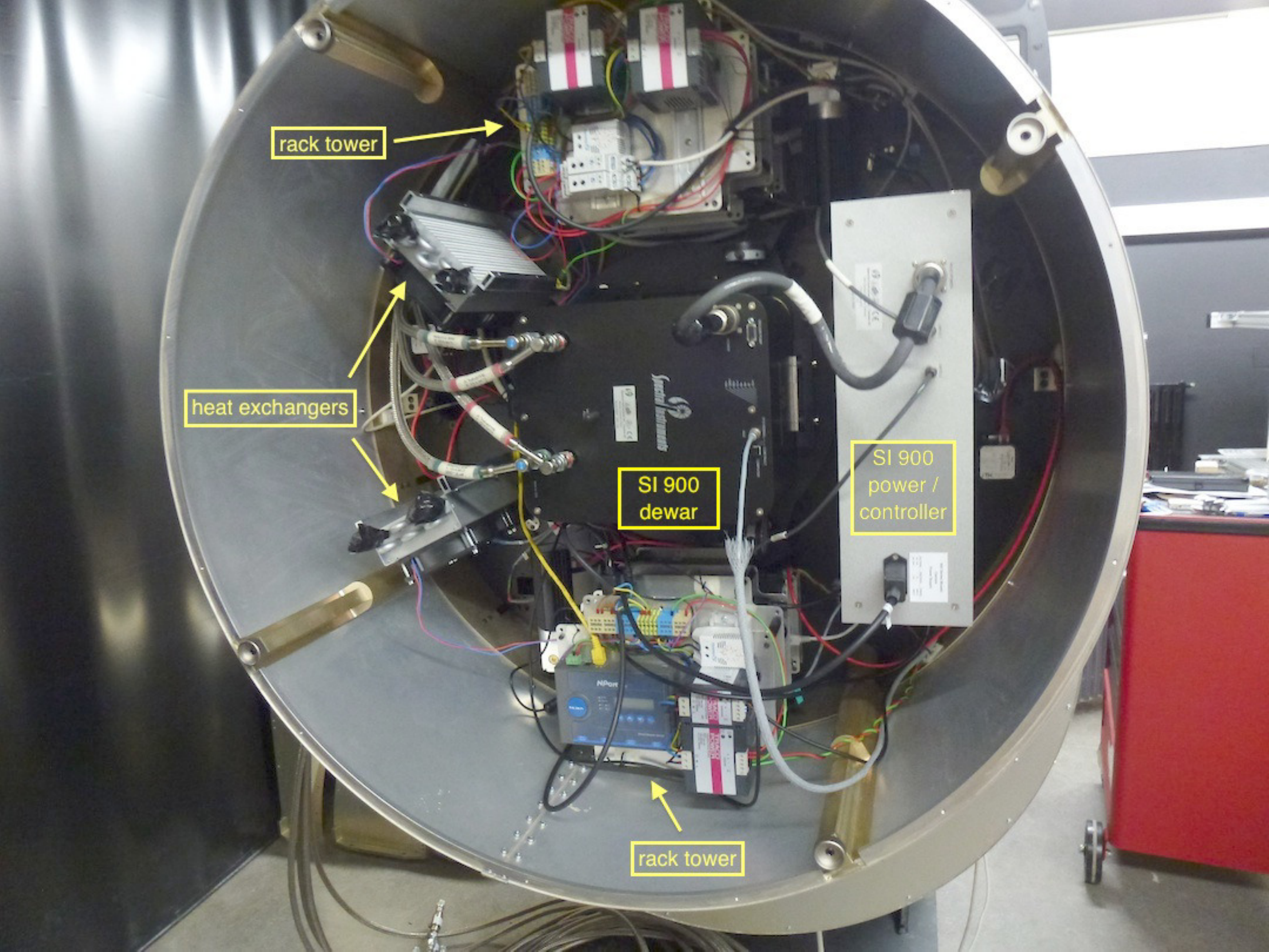}   
   \caption{WWFI mounted on the derotator test flange in the
     laboratory.
     Left: Side view with fully assembled covers. Right: Rear view
     onto the partially assembled electronics section.}
  \label{WWFI_assembled}
\end{figure*}

Fig.~\ref{WWFI_telescope} shows an image of the WWFI mounted at the
wide-field port of the {\em Fraunhofer Telescope}, and
Fig.~\ref{WWFI_assembled} shows images of the fully assembled camera
including filter wheels and EMI shielding in the laboratory and a
close-up on the WWFI cabling and electronics.
Table~\ref{WWFI_params} gives an overview of the most important camera
parameters.

\begin{table}
  \caption{Basic parameters of the Wendelstein Wide Field Imager}
  \label{WWFI_params}
  \centering
  \begin{tabular}{lc}
    \hline\noalign{\smallskip}
    \multicolumn{2}{c}{Global parameters}\\
    \noalign{\smallskip}\hline\noalign{\smallskip}
    Size (envelope) & $<1$~m radius and depth cylinder \\
    Mass & $\lesssim 350$~kg \\
    Operating temperature & $-15^{\circ}$C $\leq T \leq 25^{\circ}$C\\
    Power consumption & $\sim 1.6$~kW\\
    \noalign{\smallskip}\hline\noalign{\smallskip}
    \multicolumn{2}{c}{Optical parameters}\\
    \noalign{\smallskip}\hline\noalign{\smallskip}
    Telescope aperture & 2.0~m \\
    F-ratio & 7.8 \\
    Field of view & (27.6x29.0)~arcmin$^2$ \\             
    Pixel scale & 0.2~arcsec$/$pixel \\
    Gaps & $98"$ and $22"$ \\
    Mosaic alignment & $\leq 0.13^{\circ}$ \\
    Field distortion & $< 2.2 \cdot 10^{-5}$ \\ 
    Wavelength range & $300$~nm $\leq \lambda \leq 1050$~nm \\
    Guiding FOV &  $2 \times \sim (6.8$~arcmin$)^2$ \\
    \noalign{\smallskip}\hline\noalign{\smallskip}
    \multicolumn{2}{c}{Main detector system parameters}\\
     \noalign{\smallskip}\hline\noalign{\smallskip}
    \raisebox{-1.2ex}[0pt][0pt]{SI900 Mosaic} & $4 \times (4\mathrm{k})^2$ e2v 231-84 type \\
& deep depletion CCDs\\
    Readout time &    
    8.5~s at 500~kHz,\\
    (4 ports per CCD) & 40~s at 100~kHz\\
    \raisebox{-1.2ex}[0pt][0pt]{Readout noise} &
    $7.8~e^-$ at 500~kHz,\\              
    & $2.2~e^-$ at 100~kHz\\            
    \raisebox{-1.2ex}[0pt][0pt]{Gain} &
    $5.81 \pm 0.04~e^-/$ADU at 500~kHz,\\
    & $0.688 \pm 0.003~e^-/$ADU at 100~kHz\\
    Dark Current & \raisebox{-1.2ex}[0pt][0pt]{$0.27~e^-/$h / pix} \\
    (at $-115~^{\circ}$C) & \\
    Dynamical range & 16~bit \\
    Full well capacity & $>250$~k$e^-/$ pix \\
    Peak QE & 0.9 \\
    \noalign{\smallskip}\hline
  \end{tabular}
\end{table}

\section{Components of the WWFI}
\label{Components}

\begin{figure*}[ht]
  \centering
  \includegraphics[width=\textwidth]{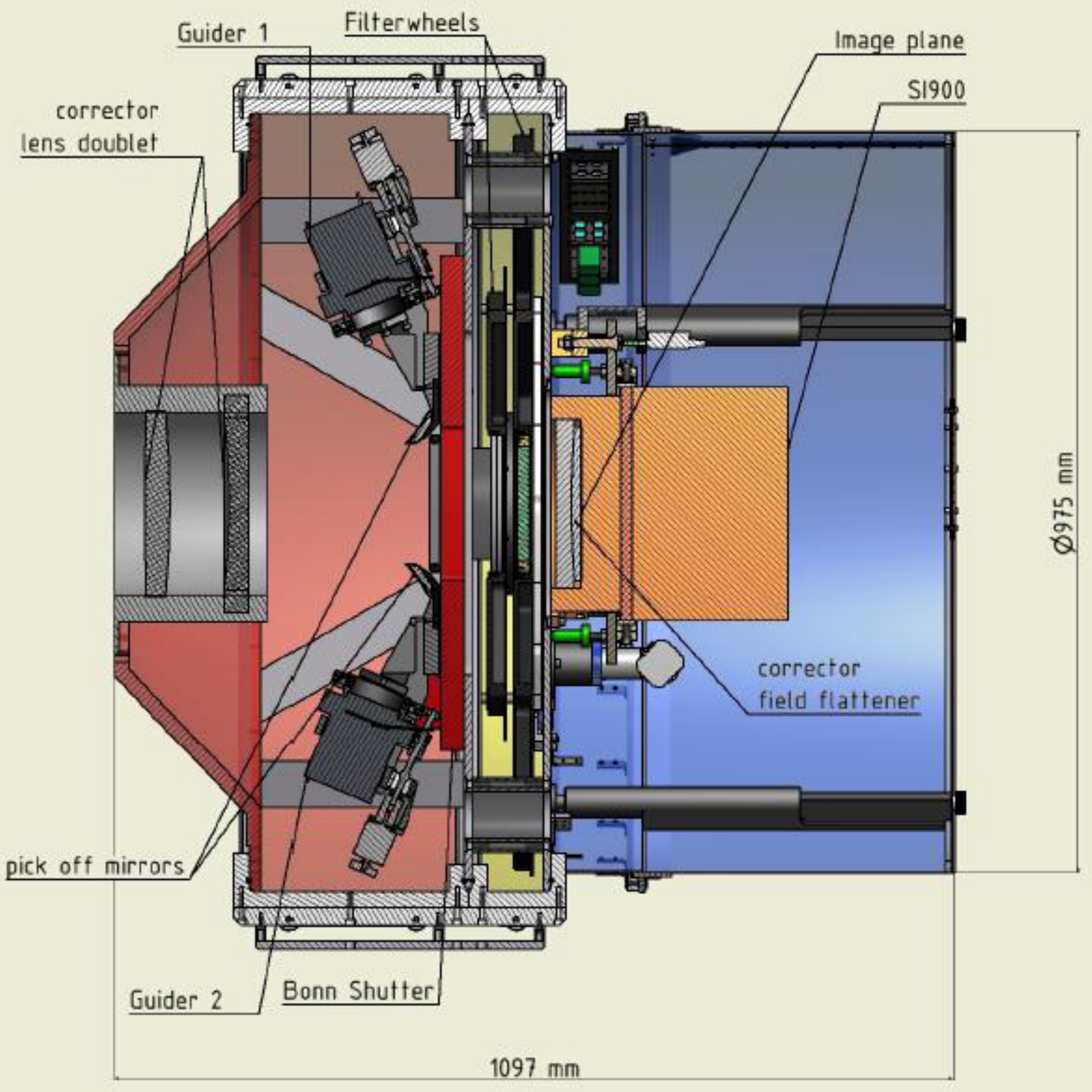}    
   \caption{Sectional view of the WWFI. Red, yellow, and blue backgrounds show its three principal sections.
   See text for details.}
  \label{WWFI_design}
\end{figure*}

\subsection{Optics and detector systems}

The WWFI is built around the wide field corrector optics, which was
integral part of the {\em Fraunhofer Telescope} package, and a Spectral
Instruments 900 series detector system (SI900\footnote{SI900 is a trademark by Spectral Instruments
Inc., Tucson, USA}).
The optical design is based on a three elements transmissive field
corrector optics and a mandatory 15~mm silica plate (or equivalent)
for filters.
The field corrector is split into a lens doublet directly attached to the
telescope flange and a field flattener lens\footnote{The field flattener was produced by POG Pr\"azisionsoptik Gera GmbH, Germany}
which also serves as entrance window of the detector dewar.
The system is designed to yield diffraction-limited images within
optical wavebands \citep{2010SPIE.7733E...5H,2010SPIE.7735E.106G}.
To map the good to excellent seeing quality of the site
\citep[$<0.8"$ median, up to $0.4"$ at best,][]{2008SPIE.7016E..58H}, a
sampling of $(0.2~$arcsec$)^2$ pixels is required which is realized by a
$2\times2$ mosaic of $(4$k$)^2$ $15~\mu$m pixel, back-illuminated e2v
CCDs\footnote{The CCDs are a trademark of e2v Inc, Chelmsford, Essex, England}.
The SI900 is a state-of-the-art scientific CCD system (see basic
parameters in table~\ref{WWFI_params} and detailed discussion in
subsections of Sect.~\ref{LabComm}).
The system employs active cooling of the mosaic with two Polycold PCC
Compact Coolers\footnote{Polycold PCC Compact Cooler is a trademark of Brooks Automation Inc, Chelmsford, USA}.
The PCC compressors are offloaded into a separate cabinet and supply
the refrigerant by 23~m long lines which run through the telescope
cable wrap. \\
The two offset guiding units pick off their light after the corrector
doublet (still in front of the main detector shutter).
While this gives partly vignetted, non-flat image planes it is easily
still good enough for guiding and allows for guide star
acquisition / guiding to be done independently from the main
shutter/filter/detector system. 
We cannibalized the CCD cameras of a previous project\footnotemark,
two Fingerlake Instruments Microline ML3041, for guiding cameras in
the WWFI.
\footnotetext{I.e.\ AMiGo, a two channel CCD-camera for the former
  80~cm telescope of the Wendelstein Observatory \citep{ediss7948}.}
Both cameras have $(2$k)$^2$, $15~\mu$m pixel, back illuminated
Fairchild CCDs 3041, use thermoelectric cooling for the detector and
had their air cooled heat sinks replaced by water cooled ones.

\subsection{Mechanics and structural analysis}

 \begin{figure}[ht]
  \centering
  \resizebox{\hsize}{!}{\includegraphics{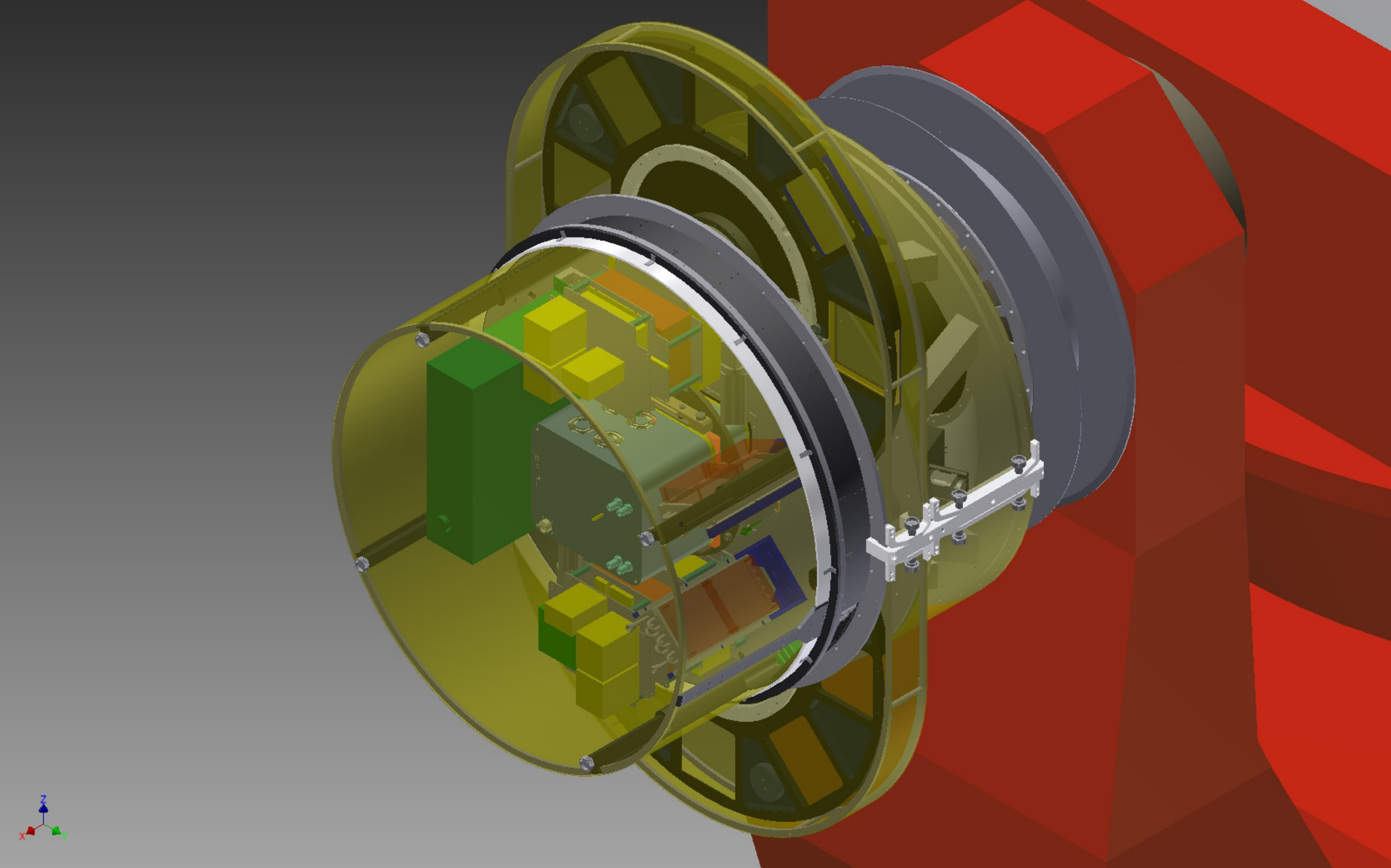}}
  \vspace*{2ex}
  \resizebox{\hsize}{!}{\includegraphics{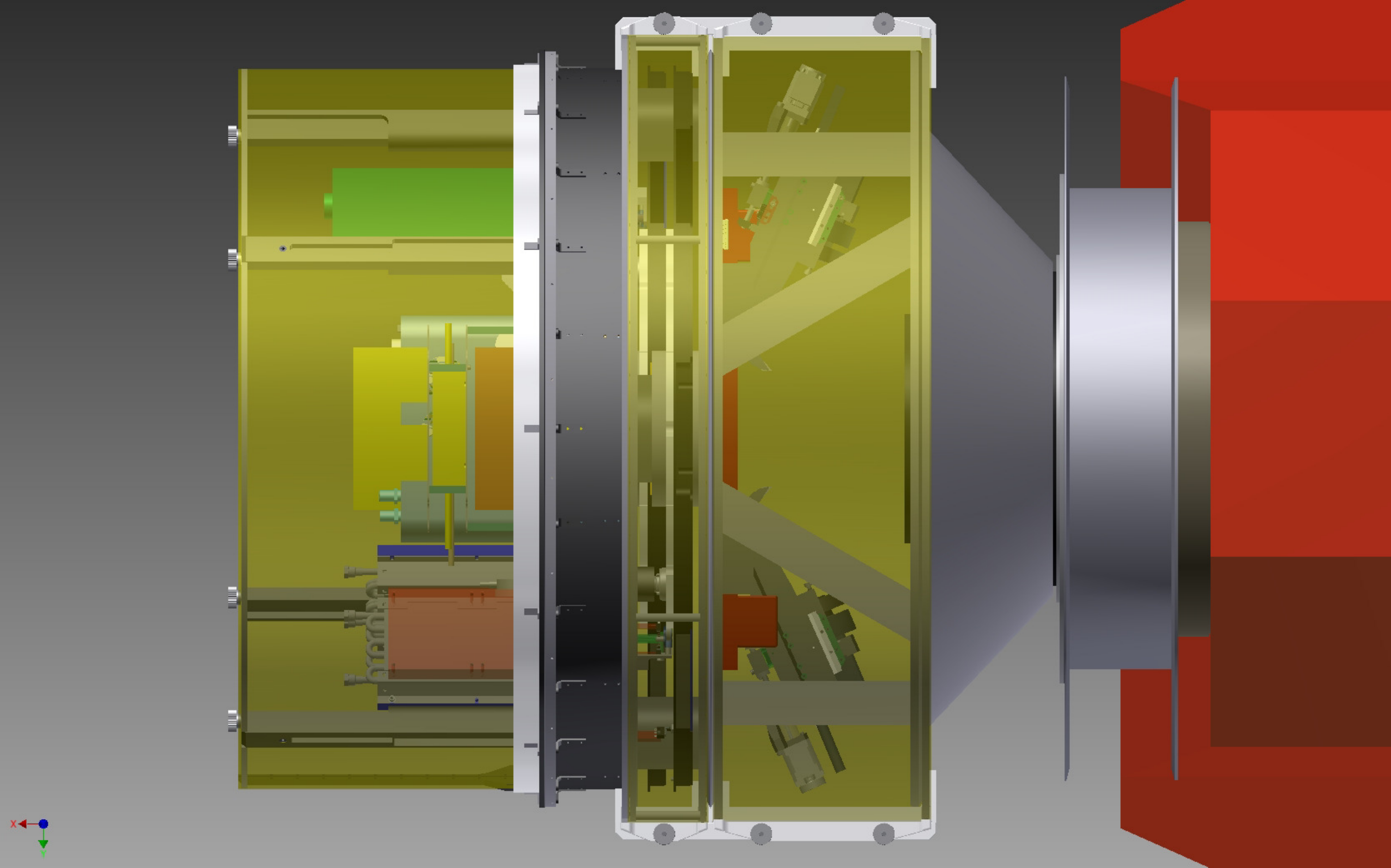} 
  \hspace*{20mm}
  \includegraphics{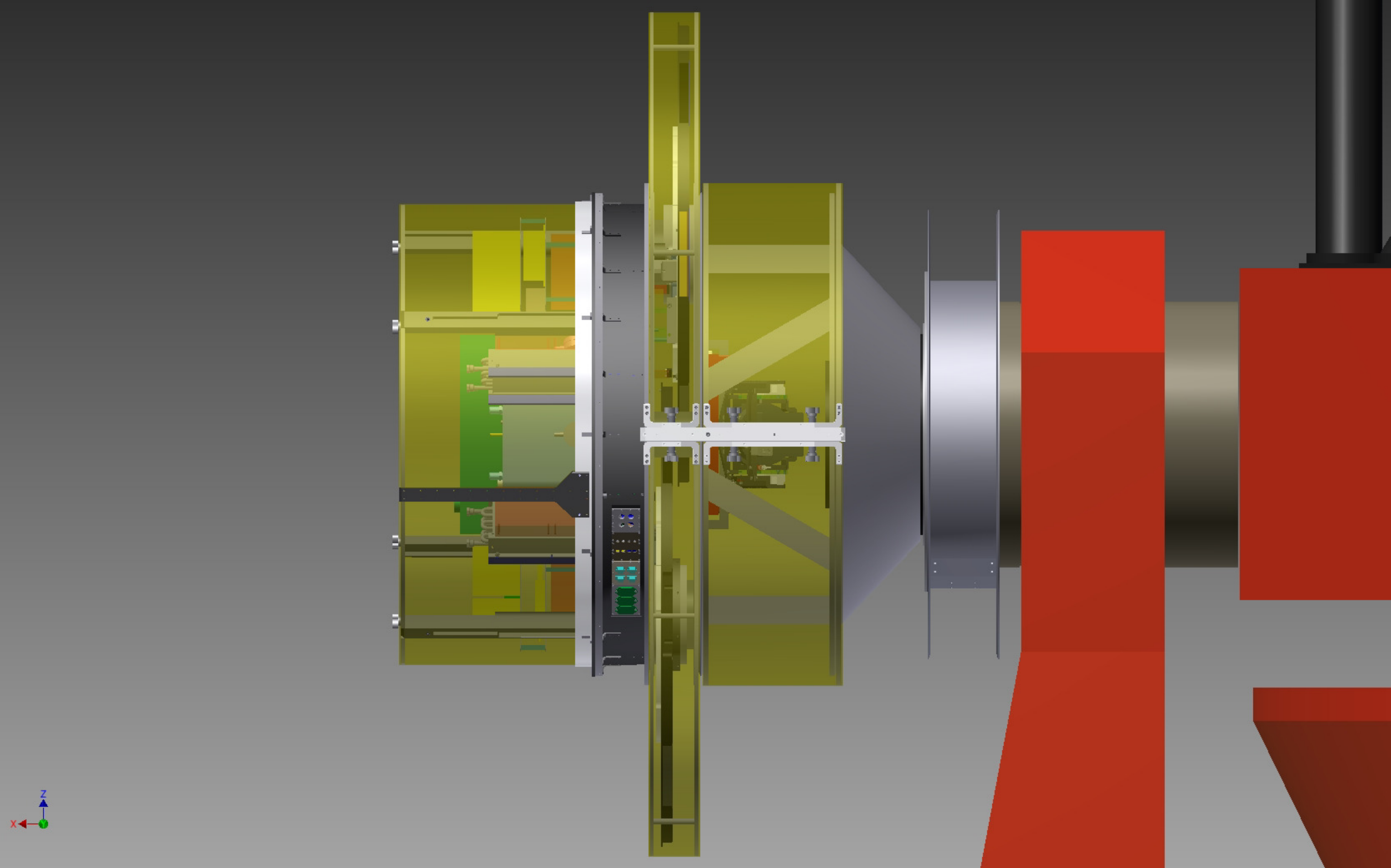}}  
   \caption{Isometric view of the WWFI (upper panel), side view
     (lower left panel) and another side view rotated by 90 degrees
     relative to middle panel (lower right panel).}
   \label{isometric}
 \end{figure}

\begin{figure}[ht]
  \centering
  \resizebox{\hsize}{!}{\includegraphics{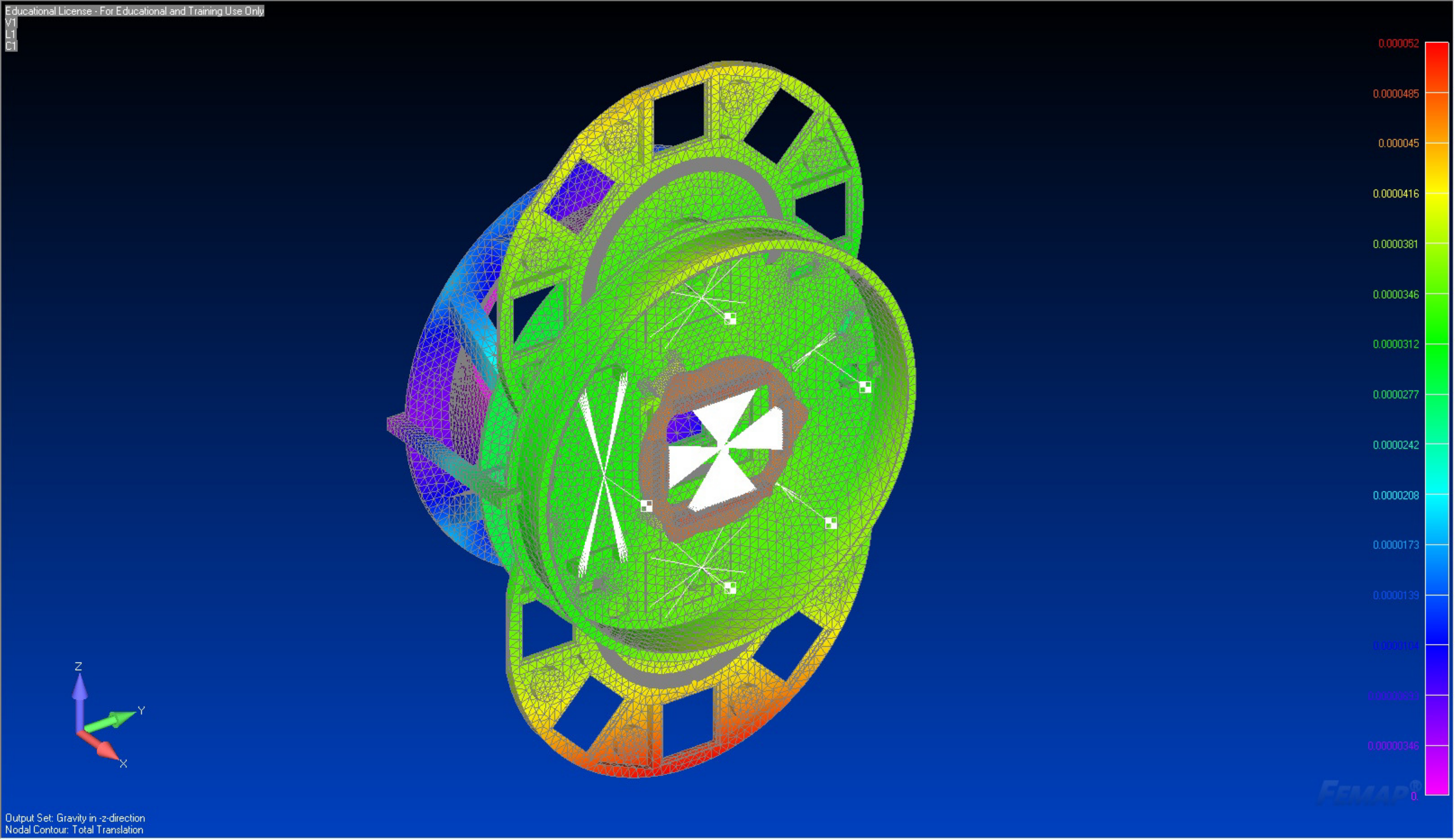}}
  \vspace*{2ex}
  \resizebox{\hsize}{!}{\includegraphics{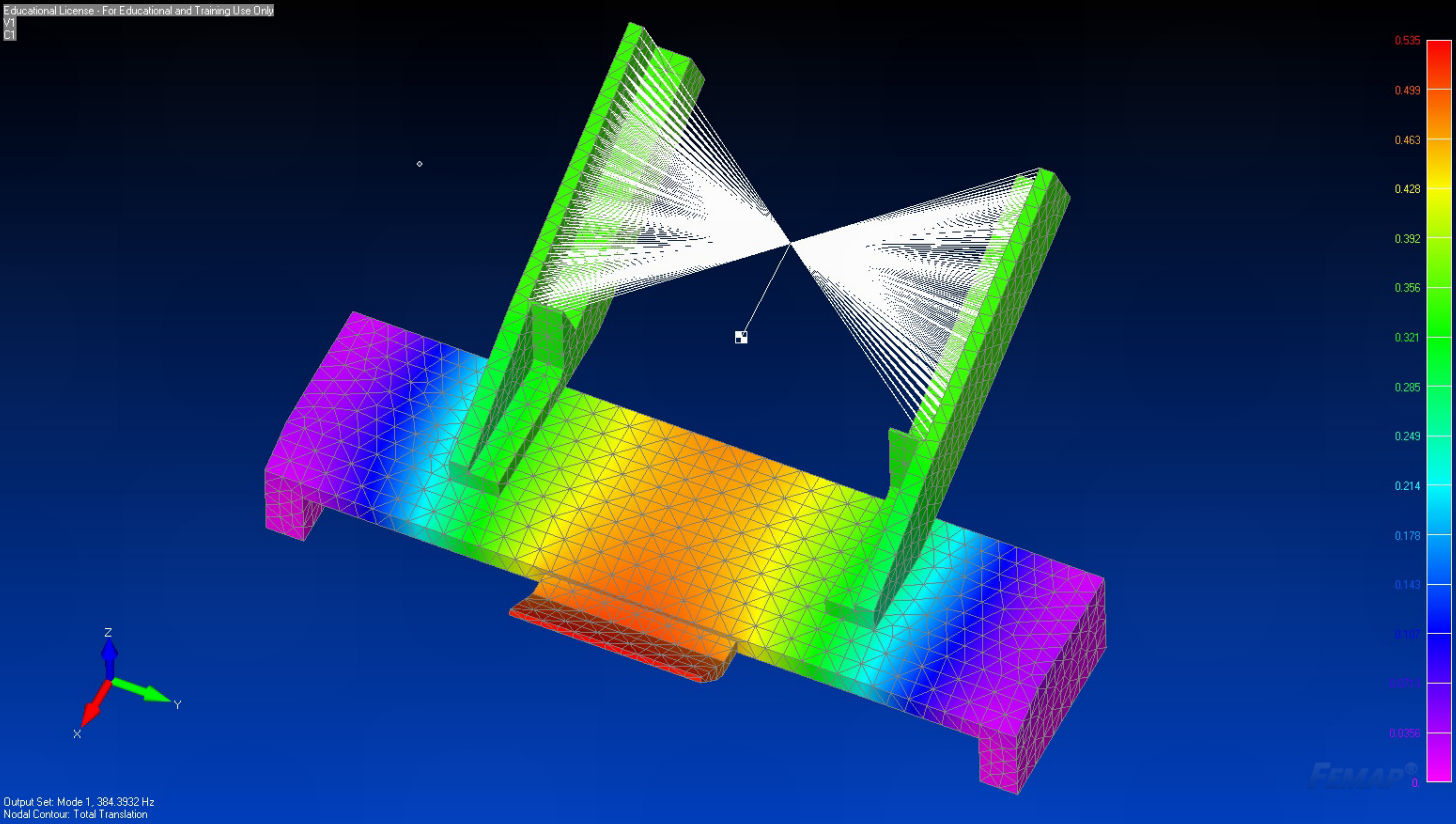}
  \hspace*{20mm}
  \includegraphics{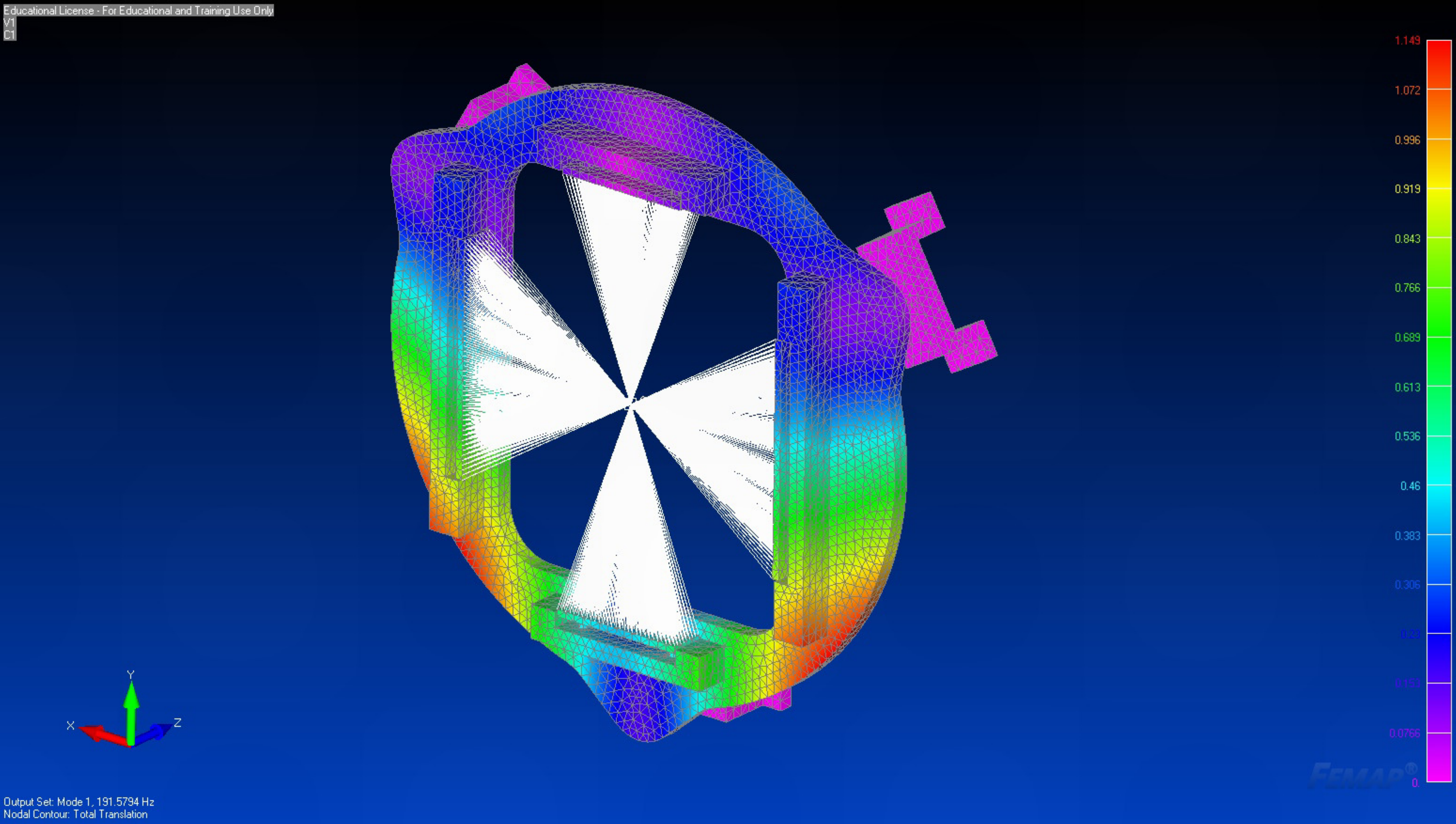}}     
   \caption{The upper panel shows an undeformated view of the total
    translation.
    Maximum displacement (red) is around 50 micron.
    The tilt angle of the parts in respect to the optical axis is
    small enough to have no influence on the optical performance.
    The lower panel shows the result of an Eigenfrequency
    analysis of the guider-mount (left) and the tip-tilt stage
    (right).
    The first Eigenfrequency of the tip-tilt stage is at 190Hz, the
    one of the guider-mount is at 380Hz.}
  \label{FEMTrans}
 \end{figure}

The mechanical design of the WWFI had to follow basic constraints
derived from the optical design and the observatory environment:
\begin{itemize}
\item It must fit inside a cylindrical volume with 1~m depth and
  radius and have its mass not exceeding 350~kg.
\item The camera has to operate at environment temperatures from
  $-15^{\circ}\mathrm{C}$ to $25^{\circ}\mathrm{C}$ without
  contributing to dome seeing.
\item The ``truss'' part of the WWFI covering the field corrector lens
  doublet has to be stiff enough to carry the whole instrument without
  significant flexure.
\item A tilt of the image plane can not be accepted while some minor
  shift of the image plane during rotation will have no discernable
  impact on image quality. 
\item Since we aim at robotic operations the WWFI should provide more
  than 10 filter slots.
\item An effective EMI protection is mandatory for any electronics to
  work due to the emissions of the close by radio transmitter.
\item Two off axis guiding cameras are needed to provide sufficient
  field and ``lever'' to correct for tracking errors of the telescope
  and its field derotator.
\end{itemize}

Our design solution aligns the corrector lens doublet, the double
off-axis guiding units, a Bonn Shutter \citep{2005AN....326Q.666R}, two
large filter wheels, and the SI900 detector system in a row
(see Fig.~\ref{WWFI_design}).
The instrument envelope is designed to act as an effective
electromagnetic interference protection against the 0.5~MW emissions
of a nearby radio transmitter. The complete mechanical design is shown
in figure \ref{isometric}.

The WWFI is divided into three sections: 
An aluminum cast cone with eight struts directly casted to it and
a $\sim 1$~m diameter mount plate enclose the first volume. 
The stiff cone covers the corrector lens doublet frame and has a small
aperture which fits to the derotator flange and a large aperture which
can sustain the rest of the camera components.
The struts form a ``Semi- Serrurier'' configuration which avoids
tilts against the optical axis of the subsequent components and are
massive enough to allow only for minor shifts perpendicular to the
optical axis (Fig.~\ref{FEMTrans}).
The telescope side of the mount plate carries the 200~mm Bonn
Shutter\footnotemark and, on top of that, two offset guiding stages.
\footnotetext{Bonn Shutters \citep{2005AN....326Q.666R} are widely
  used for large format astronomical CCD cameras, e.g.\ ESO OmegaCAM
  \citep{2006SPIE.6276E...9I}, Pan-STARRS-1 Gigapixel Camera \citep{2007AAS...211.4723T}.
Their simple and compact twin blade design yields uniform,
``photometric'' exposures even for short exposures (1~ms).}
The guiding stages each support a pick-off mirror and an FLI Microline
3041 CCD camera\footnote{FLI Microline 3041 is a trademark of Finger Lakes Instrumentation, New York, USA}
in a cardanic mount for manual tip/tilt adjustment on
a motorized linear stage\footnote{The linear stages were produced by Franke GmbH, Aalen, Germany}
for independent focusing. The linear stages are driven by stepper motors
connected to ball screws and allow for a travel range of 40~mm.
A precise MYCOM limit switch\footnote{The precision switches were produced by MYCOM AG, Berlin, Germany}
is used as reference for initialization;
counting motor steps gives the relative position.

The second volume holds two eight-position filter wheels in between the
guider/shutter mount plate and a second mount plate for the SI900 detector
system and the electronics.
There are 14 slots for filters as one empty slot is needed in each                 
wheel.
The first wheel (next to the science camera) is already equipped with an SDSS
filter set \citep[$ugriz$,][]{1996AJ....111.1748F}.
The size of the filters is $(150~$mm$)^2$ in the first and
$(160~$mm$)^2$ in the second filter wheel
(as the distance from the second wheel to the focal plane is slightly
larger).
For now we have also installed a black metal sheet filter in each
wheel to allow for additional stray light and EMI tests.
The plates are attached to each other with four short thick ``tubes''.
Two of these contain shafts for the bearings that hold the
wheels, all four can be used to feed support lines from the last
section through to the first.
For a repeatable positioning of the filters we employ a notch
mechanism.
The wheels are driven by stepper motors attached to a gearbox with a
gear ratio of 12:1 which provides the torque needed to drive the system.
We installed two ``limit'' switches to get information about the
position of the notch itself (notch in the groove or not) and one
extra switch to define a reference position\footnotemark.
\footnotetext{See next Sect.\ for details on drive logics.}

The backside of the second plate carries the camera head and all
electronics\footnotemark
\footnotetext{I.e.\ power supplies, RS232 to Ethernet converters,
thermostats, switches, motor controllers, compressor relays, and
embedded control PCs.}
needed to drive and control the WWFI components.
The back focus tolerance of the telescope optical design was $\pm
  \mathrm{4mm}$.
Therefore, and to allow for less tight tolerances when machining the
mechanical parts, we mounted the camera head with a manual 5-axis
(tip/tilt and x, y, z translation) stage onto the plate.
This electronics volume is insulated with Armaflex\footnote{Armaflex is a trademark of Armacell GmbH, M\"unster, Germany}
and cooled by two
liquid-to-air heat exchangers from Thermatron Engineering Inc.\ to
minimize the contribution to dome seeing of the instrument.

We also did a finite element method (FEM) analysis of the WWFI to have a look at the
Eigenfrequencies and bending behavior.
Because all telescope axes (azimuth, elevation and both derotators)
are driven by direct drives we had to make sure that the
Eigenfrequencies of the structural parts are high enough ($>50$~Hz),
to lower the risk of mechanical oscillations induced by the direct
drive controllers.
Because of the complexity of the model, the FEM analysis was split
into several steps.
First we had a look at subassemblies as the guider mechanism or the
heat exchanger mounts.
When the FEM model showed that the Eigenfrequencies are high enough, we
integrated the part only as mass point in the complete FEM model of
the WWFI.
This helped to keep the complexity small enough to have reasonable
calculation times.
Only the electronics mounts and one sheet metal of the EMI housing
that covers the electronics and camera head showed Eigenfrequencies
low enough to possibly get excited.
The sheet metal was damped by the Armaflex insulation we attached to it.
We haven't encountered any problems yet with this part.
We also put some Armaflex insulation beneath the electronics mounts to
have a soft connection to the stiff structure.
The lowest Eigenfrequency of the supporting structure was found at 83 Hz 
(see Fig.~\ref{FEMTrans}).

The other value we were interested in was the bending behavior.
For this we used the same FEM model as for the Eigenfrequencies,
because everything necessary was already implemented there (mass
points for subassemblies, connections, mesh). The only thing that
needed to be done was switching on gravity in different directions to
get the bending behavior.
We were especially interested in the differential bending between the
detector surface and the guider, because of its impact on guiding performance.
The differential translation we found was negligibly small.
The maximum total translation at the camera surface was around
$50~\mu$m (see Fig. \ref{FEMTrans}).

\subsection{EMI covers}

The covers of the WWFI not only serve as a shield from light but also
from EMI (electromagnetic interference) due to a close by radio
transmitter station.
The camera has to work within fields $\approx 20\,$V/m.
Without an effective shield the detector displays enhanced noise
(Sect.~\ref{sectNoise}) and the motor controllers for the filter wheels 
and offset guider focus movement just do not work at all.
(They pick up too much interference from the lines to the
limit/position switches to boot properly.)
The 5-part cover is built from chromated aluminum sheets
screwed and conductively glued onto a minimal truss.
High conductivity glues have about 80\% filling of silver (or a similar
conductive metal) and therefore are not adhesive enough without the 
additional screws to hold the sheets in place.
The ``sharp'' edges of the covers slide into light traps with
conductive lip seals.
The only electric lines into the camera are shielded and filtered
power lines; network connection is established via optical fiber
link.
Hierarchized thermal switches protect the electronics from overheating
in case of a cooling failure.

\subsection{Software and control}
The WWFI control software has to support and combine the different
proprietary interfaces of its hardware components:
The SI900 is controlled through a Windows graphical user interface (GUI, 
based on LabView\footnote{LabView is a trademark of National Instruments Corporation, Austin, USA})
which offers a TCP/IP socket for ``backdoor'' control.
The FLI MicroLine 3041 guiding cameras come with a C Developer Kit for
Linux.
The filter wheels and offset guiding focus work through Pollux\footnote{Pollux Controller and Venus-2
command language are trademarks of PI miCos GmbH, Eschbach, Germany} high
resolution positioning controllers via the Venus-2 command language on
serial interfaces (which we map to TCP/IP via Moxa NPort\footnote{Moxa NPort is a trademark
of Moxa Inc., Brea, USA}).
The Bonn shutter is directly controlled by an I/O signal from the
camera but also offers additional controlling and surveillance options
through a serial interface of its motor controller (again mapped to TCP/IP).
For all four components we developed device programs which can be
accessed by TCP/IP sockets and translate simple human readable
commands to the explicit hardware control commands and vice versa for
the messages received from the hardware.
The device programs log state and optionally debug messages to a
central syslog facility server which again parses a subset of those
messages to provide status webpages (simple HTML) which are
independent of the higher level controlling software.
They also already allow for ``scripted'' observations which
greatly enhance the efficiency of commissioning.

While the device programs were planned to map only basic functions of
their respective hardware there had to be some exceptions to that
rule:
The motor controller of the filter wheels and its language was
specifically designed for arbitrary linear movements between hard
limits which is obviously almost the opposite of moving between
mechanically fixed positions on a circle.
Therefore, we use the position switch as a simultaneous upper/lower
limit switch with the reference switch inverting the upper limit
again\footnotemark.
\footnotetext{Because of this the initialization run has to move
  ``backwards''.}
Now, as the switches reset the position accounting within the
motor controller the device program has to count filter notches.
It also has to turn off hard limits before starting moves and turn
them back on while moving as active limit switches control the
direction in which subsequent moves are allowed.
The second exception is guiding image evaluation.
As the device program already holds the images (before optionally
saving them to disk) it is also the right place to evaluate them,
i.e.\ to correct for bias / dark current, compute star positions
and perform a rudimentary point spread function (PSF) analysis (second order moments).
This saves bandwidth and improves performance (speeds up guiding turn
around) as the higher level control instance runs on another platform.

The next layer of software represents the logically integrated WWFI
control:
It connects to the single device programs and again offers simple
human-readable commands and messages on its TCP/IP interface to
control the instrument.
It allows to start / stop guiding, move filters, expose etc. while
keeping track of the individual components and prohibits ``stupid''
mistakes (like changing filter while exposing).
This layer now can not only be controlled from the command line but
also via a web-browser based GUI or a robotic scheduler.
Our prototype for this layer which already provides the guiding for the
WWFI makes heavy use of multithreading and is implemented in Python\footnote{Python
Programming Language is a trademark of Python Software Foundation, Beaverton, USA}
(work in progress).

\section{Calibration and commissioning}
\label{LabComm}

In this section we describe all laboratory measurements that we
have performed with the WWFI as well as the on-sky calibration and
present the results.
The tests include gain, quantum efficiency, charge
transfer efficiency and charge persistence measurements as well as
photometric zero point calibration and an on-sky calibration with
stellar spectra.

\subsection{Gain}
\label{sectGain}

\begin{figure}
   \centering
   \resizebox{\hsize}{!}{\includegraphics[angle=0]{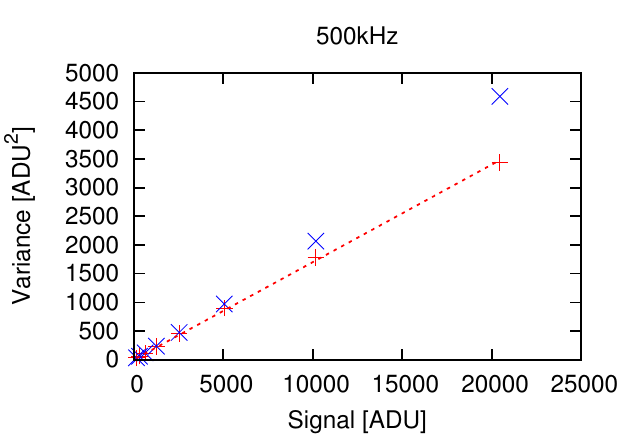}
   \includegraphics[angle=0]{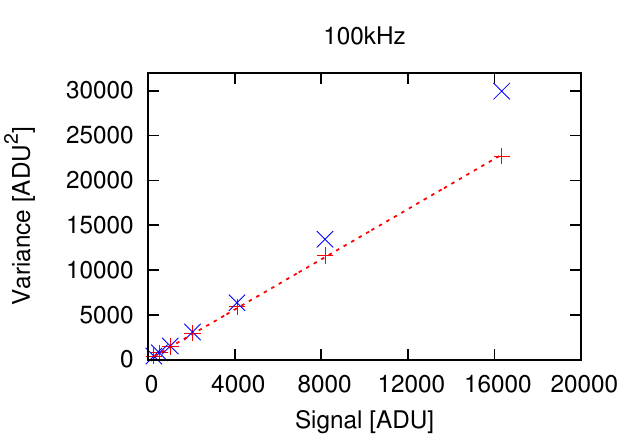}}
   \caption{Exemplary photon transfer for the 500~kHz readout mode
     (left) and the 100~kHz mode (right) for CCD 0, Port 1,
     with the signal in ADU on the x-axis and the variance on the y-axis.
     The blue ``$\times$'' show the uncorrected values, while the red ``+'' show
     the values that have been corrected for the noise of the masterflat.
     An early version of this figure is shown in \cite{2012SPIE.8446E..3PG}.}
   \label{PhotTrans}
 \end{figure}

\begin{table}
 \caption{Gain per port for the fast and slow readout mode (500 and
   100~kHz), as measured in our lab.}
  \label{Gain_tab_fastslow}  
\centering
\begin{tabular}{ccccc}
\hline\noalign{\smallskip}
 & \multicolumn{4}{c}{Gain [$e^- / $ADU]} \\
\noalign{\smallskip}\hline\noalign{\smallskip}
CCD & 0 & 1 & 2 & 3 \\
  & 500 / 100 & 500 / 100 & 500 / 100 & 500 / 100  \\
Port & [kHz] & [kHz] & [kHz] & [kHz] \\
\noalign{\smallskip}\hline\noalign{\smallskip}
1 & 5.87 / 0.71 & 5.94 / 0.71 & 5.87 / 0.71 & 5.85 / 0.71 \\
2 & 5.88 / 0.69 & 5.85 / 0.68 & 5.84 / 0.70 & 5.87 / 0.69 \\
3 & 5.76 / 0.68 & 5.75 / 0.67 & 5.73 / 0.67 & 5.72 / 0.67 \\
4 & 5.75 / 0.68 & 5.78 / 0.68 & 5.75 / 0.68 & 5.79 / 0.69 \\
\noalign{\smallskip}\hline
\end{tabular}
\end{table}

The {\em gain} of a photon collecting device is given by the ratio
\begin{equation}
 g = \frac{N_e}{\#\mathrm{ADU}}
 \label{gain_eq}
\end{equation}

To measure the gain factor we made use of the {\em photon transfer
  gain method}, as it is described in \cite{mclean2008electronic}:
In principle it would be necessary to take multiple flat field
images at multiple illumination levels and measure the mean signal and
noise for each and every pixel on the chip at each illumination level.
Instead of analyzing each pixel in several (equally illuminated) flat
field images, we take only one image per illumination level and
substitute the averaging over several images by averaging over several
pixels and previously removing the pixel-to-pixel variations by
dividing each image by a {\em masterflat} composed of 30 single
flat-field images at a signal level significantly below half well
capacity.
Then we determine the mean signal and variance of every image (one per
illumination level).
For further considerations we can neglect the readout noise as
it is well below the photon noise, so the photon noise $\sigma$ is the
only source of variance $\sigma^2$ left in an image with an average
signal $S$, since $noise^2=p^2 + R^2$ (with photon noise $p$ and readout noise $R$).
Dividing this equation by the squared gain, the left hand side becomes the variance (in ADU) and 
since the photon noise p is equal to the square root of the signal $\sqrt{g \cdot S}$, we get:

\begin{equation}
 \sigma^2 = \frac{S}{g}
\label{Noise}
\end{equation}

Unfortunately, the introduction of the masterflat also introduces
additional photon noise. We used the method introduced in
\cite{2012SPIE.8446E..3PG} to correct for this additional noise, the following
description closely follows the derivation therein. 

The relative noise in the final signal ($F_i$) is given by:

\begin{equation}
 \left (\frac{\sigma_{F_i}}{F_i} \right )^2 = \left (\frac{\sigma_M}{M} \right )^2 + \left (\frac{\sigma_{S_i}}{S_i} \right )^2
\label{PhotonNoise}
\end{equation}

where $S_i$ is the average signal in the original exposure (index $i$ for
number of the exposure), $M$ is the average signal of the masterflat and
$F_i$ the average signal in the final image (divided by the
masterflat) and the $\sigma$ the corresponding photon noises.
Since the masterflat is normalized to 1 we can assume that $S_i = F_i$ and use 
equation \ref{Noise} to obtain the following equation for the gain:

\begin{equation}
 g = \frac{ \frac{1}{F_i} - \frac{1}{F_j}}{\left ( \frac{\sigma_{F_i}}{F_i} \right )^2 - \left (\frac{\sigma_{F_j}}{F_j} \right )^2}
\label{gain}
\end{equation}

for any indices $i \neq j$, for all pairs of data points. 
The gain is now estimated via equation~\ref{gain}, to determine the (relative)
photon noise in the masterflat via equation~\ref{Noise}, which is then 
subtracted in in equation~\ref{PhotonNoise} to obtain the pure photon noise, corrected for 
the contribution of the masterflat.
Figure~\ref{PhotTrans} shows the photon transfer functions for the
500~kHz (top) and 100 kHz (bottom) readout mode,
with blue ``$\times$'' for uncorrected values and red ``+'' for
values corrected for the noise of the masterflat.
The gain has finally been determined as the slope of the linear fit to
the corrected values.
Table~\ref{Gain_tab_fastslow} shows the gain for both readout modes
for all ports and CCDs.

\subsubsection{Relative gain calibration}
While the absolute gain determination is not better than a few \% we
used flat-fields to adjust the gains within one detector to be
consistent to each other to better than 0.05\%.
Usually flatfielding would take care of those minor differences as a
per port individual multiplicative gain factor is applied to both flat-field and
science images and therefore cancels out. But adjusting gain levels
helps us overcome differential bias level fluctuations at the
0.3$e^-$ level.
We use clipped averages of almost adjacent rows/columns\footnotemark \
for correction factors.
\footnotetext{We used the third row/column next to the border.}
The ``almost'' is because the CTE (see Sect.~\ref{sectCTE}) is
affecting the last read out rows/columns enough to give overall wrong
correction factors if those were used.

\subsection{Bias level calibration}
As mentioned before the bias level and even its offset between serial
overscan\footnotemark \ and the image region is not stable to more than
about 0.3$e^-$.
\footnotetext{We read three overscan regions from each port: Serial
  pre-  and overscan, as well as parallel overscan. The serial overscan
  displays the smallest and most stable offset to the image region in
  bias and dark frames.}
The resulting ``small'' steps between different ports within one
detector can yield rather large distortions of the isophote shapes of
extended objects (galaxies) which fill more than one quadrant of a
detector.
Since we have calibrated the gain ratios within one detector we can
apply the same principle again for scientific images with big enough
regions of low flux levels at the port boundaries.
(For medium to higher flux levels $\sqrt{\mathrm{flux}[e^-]} \gg 0.3\,e^-$ the steps
are irrelevant).
Here we use median clipped averages of the directly adjacent
rows/columns to derive and correct for the remaining bias offsets
between the detector ports.

\subsection{Readout noise}
\label{sectNoise}

There are three types of noise present in CCD images: readout noise, photon noise and
pixel noise. A detailed description of the noise types in a CCD can be found in 
\citet{janesick2001scientific}.

\begin{table}
\caption{Average gain and readout noise measured in the lab (USM), by
  the manufacturer (SI) and typical values measured at Mt.\ Wendelstein without EMI-shield (WST)
  and with EMI covers (WST-shield).
  The values of the readout noise show clearly that the presence of the radiation raises the noise drastically
  (by about 50\%), but the EMI-shield mitigates this effect (for slow
  readout even completely). The readout noise varies less than 0.2 ADU for lab and EMI protected frames, 
  but can change for several ADU between different not EMI protected frames on site.}
\label{ReadoutNoise}  
\centering
\begin{tabular}{ccccc}
  \hline\noalign{\smallskip}
  Readout & \multicolumn{4}{c}{Gain [$e^-$/ADU]} \\
  mode& \multicolumn{2}{c}{USM} & \multicolumn{2}{c}{SI} \\
  \noalign{\smallskip}\hline\noalign{\smallskip}
  500~kHz & \multicolumn{2}{c}{$5.81  \pm 0.04$}  & \multicolumn{2}{c}{5.89} \\
  100~kHz & \multicolumn{2}{c}{$0.688 \pm 0.003$} & \multicolumn{2}{c}{0.72} \\
  \hline \hline
  Readout & \multicolumn{4}{c}{Noise [$e^-$]} \\
  mode & WST & WST-shield & USM &  SI \\
  \noalign{\smallskip}\hline\noalign{\smallskip}
  500~kHz &  12.4  & 8.0 & 7.8 & 8.1 \\
  100~kHz &  3.3 & 2.2 & 2.2 & 2.4 \\
  \noalign{\smallskip}\hline
\end{tabular}
\end{table}

The readout noise can be determined by measuring the noise of a bias frame.
We did this for all 16 ports of the WWFI by determining the noise of a whole image 
and clipping of outliers above $5~\sigma$ in order to remove defective pixels.
Table~\ref{ReadoutNoise} shows the average values of the readout noise
for both readout modes measured in our lab compared to the results of
the manufacturer.

The values measured in our lab are systematically lower than the ones
achieved by SI.
The reason for this are the slightly lower gain values we measured.

We also checked the noise difference in the laboratory and on-site
with and without the electromagnetic shielding.
The results show that the noise on-site is about $50\%$ higher
due to the strong radiation and the shield mitigates this effect
(for slow readout even completely).
Additionally, we checked the contribution of the charge quantization
to the readout noise: we found no difference in the fast readout mode,
while in the slow mode we measured a quantization noise of 0.02
electrons, which is negligible for all our applications.

\subsection{Quantum efficiency}
\label{sectQE}

The quantum efficiency (QE) of a detector is the fraction of photons
incident on the surface of the device that produce charge carriers.
It is measured in terms of electrons per photon and is a function of wavelength.

Next, we describe our method to measure the QE in the laboratory and
compare our results for all four chips  with the results obtained by
the CCD manufacturer e2V.

\subsubsection{The setup}

\begin{figure*}
  \centering
  \includegraphics[width=\textwidth]{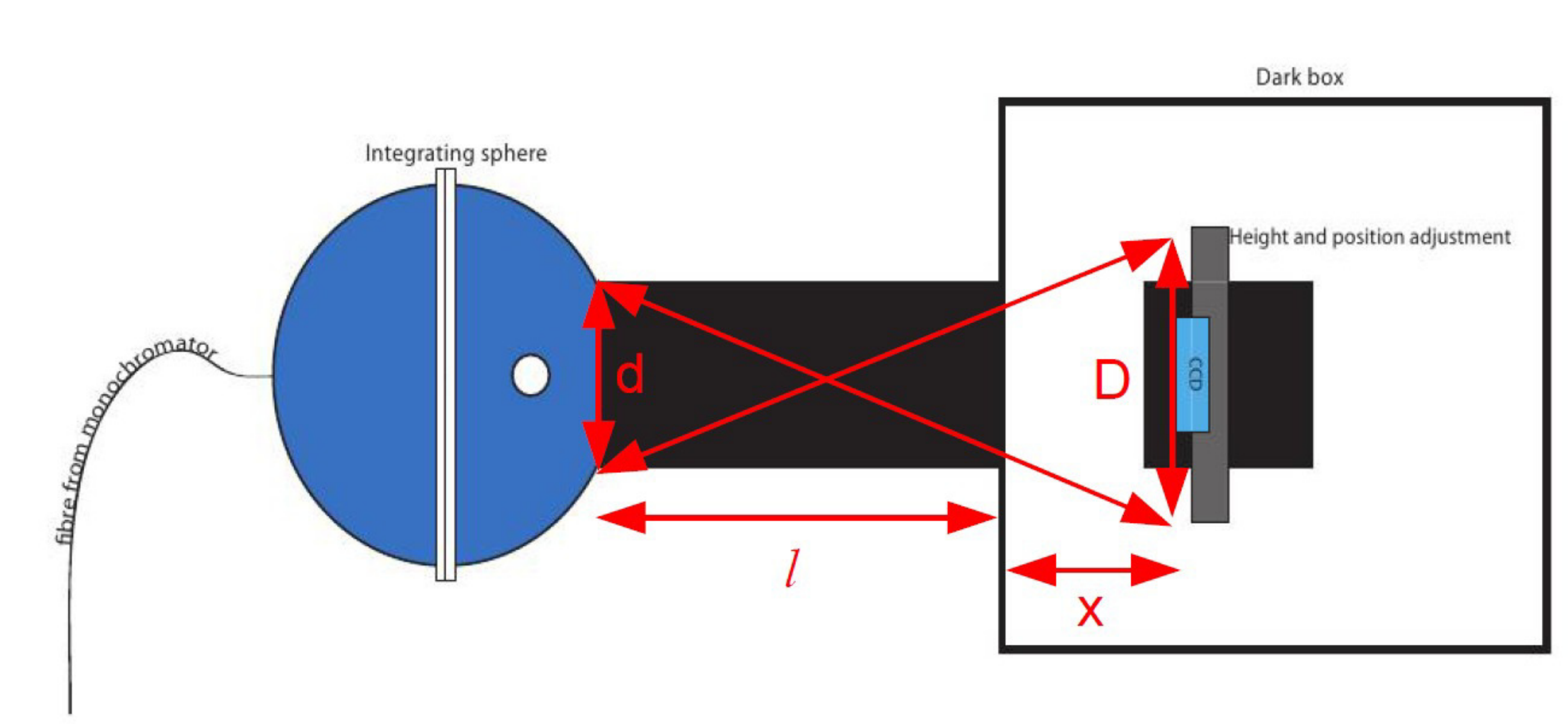} 
  \caption{Sketch of test setup with integrating sphere and darkbox}
  \label{Darkbox}
\end{figure*}

Measuring the QE of a detector requires a homogeneously illuminated
area at least as large as the collecting area of the detector, 
which in our case is $\sim~(15~\mathrm{cm})^2$.
We used a 100 W white halogen lamp as source of illumination.
After passing through a double-monochromator for wavelength selection,
the monochromatic light is fed to an integrating sphere (via an optical
fiber) which randomizes the direction of the light rays and creates a uniformly
illuminated source.
The flat light from the sphere passes through a tube with a
diameter of 30~cm to a large darkbox where the detector is mounted at
a distance that corresponds to a focal ratio of $f/7.8$, which is the
same as at the Fraunhofer telescope in order to simulate the incident angles 
as they are at the telescope site.
A calibrated photodiode was used to measure the absolute amount of
photons per unit area arriving at the camera plane in the dark box.
Figure~\ref{Darkbox} shows a sketch of the integrating sphere and the
darkbox.

\subsubsection{Measurement}

The quantum efficiency of the camera was measured in the
wavelength region 340 - 1000~nm, in 20~nm steps up to 900~nm and in 50~nm
steps above 900~nm.
At each wavelength, five images were taken with an exposure time
just high enough that the average amount of counts is something around
10000 ADU.
Additionally, a single dark frame was taken for each exposure time.

\subsubsection{Data analysis}

The definition of the gain is given in equation \ref{gain_eq} 
and the definition of the quantum efficiency of a detector is given by:

\begin{equation}
\mathrm{QE} = \dfrac{N_e}{N_\mathrm{phot}} 
\end{equation}

with $N_\mathrm{phot} = \dfrac{P \cdot t_\mathrm{exp} \cdot \lambda}{h \cdot c}$ (where $P$ is the 
power of the incident light $P=\frac{dE_{phot}}{dt}$ and $I$ is the photodiode current) and
the spectral response of the photodiode
$\mathrm{SR}:=\dfrac{I}{P} \Rightarrow P = \dfrac{I}{\mathrm{SR}}$ and the transmissivity
of the entrance window $T_\mathrm{win}$ we obtain:

\begin{equation}
\mathrm{QE} = \dfrac{g \cdot \#\mathrm{ADU} \cdot \mathrm{SR} \cdot h \cdot c}{I \cdot t_\mathrm{exp} \cdot \lambda \cdot T_\mathrm{win}}
\end{equation}

Since the detection area of the photodiode ($A_{pd}$) is different from the
active area of a CCD-pixel ($A_{pix}$) we need to multiply the equation by the 
ratio of the two areas:

\begin{equation}
\mathrm{QE} = \dfrac{g \cdot \#\mathrm{ADU} \cdot SR \cdot h \cdot c}{I \cdot t_{exp} \cdot \lambda \cdot T_{win}} \cdot \dfrac{A_{pd}}{A_{pix}}
\end{equation}

Table~\ref{QE_quantities} explains all parameters and quantities used in this derivation.
There are two problems arising in our setup concerning the reference
measurement with the photodiode:
First, the measurements with the CCD and the diode should take place
simultaneously, or to be more exact, the time interval between the
measurements must be shorter than the time in which the illumination
from the lamp changes significantly.
The current-stabilized power supply of the halogen lamp provides constant illumination
over a time period of a few hours, so the time interval between
the measurements should be much less than that, which
cannot be realized in our setup.
Second, the amount of light incident at the camera plane is very low.
At short wavelengths (where the spectral response of the photodiode is low) it
is therefore not possible to measure a signal with the diode at all.
At the surface of the integrating sphere however, the illumination is higher by
approximately a factor 100.
We solved these problems by measuring the diode current at the surface
of the integrating sphere simultaneously with the CCD measurement.
With this measurement we only determine the number of photons per unit
area at the surface of the integrating sphere, but we need to know it
in the plane of the camera.
Now we need to measure the ratio of the illuminations in the sphere
and in the camera plane, which can be done in a separate measurement,
where the two values for the diode current can be taken within few
minutes and thus one does not have any problem with the instability of the light
source.
With these two values for the diode current we generated a calibration
factor that is equal to the ratio of the illuminations in the sphere
and at the camera plane:
$c_f = \dfrac{L_\mathrm{sphere}}{L_\mathrm{camera}}$

The illumination ratio can also be estimated by geometrical
considerations:
Let $d$ be the diameter of the tube through which the light leaves the
integration sphere, $D$ denotes the diameter of the illuminated area
in the camera plane, $l$ is the length of the tube and $x$ is the
distance of the camera from the front wall of the dark box (see red
lines and arrows in Fig.~\ref{Darkbox}).
All of these quantities can be measured directly except for $D$
which can be calculated:
$\frac{D}{l/2+x}=\frac{d}{l/2}$ or $\frac{D}{d}=1+\frac{x}{l/2}$.
The illumination on the surface of the integrating sphere is
proportional to $\frac{1}{d^2}$ while the illumination in the camera
plane is proportional to $\frac{1}{D^2}$, so the ratio of
illuminations is equal to $\frac{D^2}{d^2}$.
With the numbers from our setup $x = 89~cm$, $d = 30~cm$ and
$l = 80~cm$ we get an illumination ratio of $10.4$.
The (wavelength-averaged) illumination ratio from our measurement is
$33$ which means that we lose more than a factor 3 more light than we
expect from our (simple) estimation.
Remembering that inside of the sphere all angles of light rays are
present, while in the camera plane there are only light rays under
steep angles given by the geometry (the flat angles hit the inside of
the tube which is black and will absorb the most), we can instantly
say that our simple approximation underestimates the illumination
ratio by an amount which is given by the geometry (i.e.\ the minimum
acceptance angle of light rays incident at the camera plane).
This ratio enters the equation for the QE as a linear factor:

\begin{equation}
\mathrm{QE} = \dfrac{g \cdot \#\mathrm{ADU} \cdot \mathrm{SR} \cdot h
  \cdot c}{I \cdot t_\mathrm{exp} \cdot \lambda \cdot T_\mathrm{win}}
\cdot \dfrac{A_\mathrm{pd}}{A_\mathrm{pix}} \cdot c_f 
\end{equation}

\begin{table}[ht]
\caption{Quantities used in QE equation.}
\label{QE_quantities}
\centering
\begin{tabular}{ll}
\hline\noalign{\smallskip}
QE & quantum efficiency of CCD \\
$g$ & gain (ratio of electrons per ADU) \\
\#ADU & number of analog to digital counts \\
$SR$ & spectral response of the photodiode in $\dfrac{A}{W}$ \\
$h$ & Planck's constant \\
$c$ & speed of light \\
$I$ & current of the photo diode \\
$t_\mathrm{exp}$ & exposure time of the image \\
$\lambda$ & wavelength \\
$A_\mathrm{pix}$ & area of one pixel \\
$A_\mathrm{pd}$ & area of the photodiode \\
\raisebox{-1.2ex}[0pt][0pt]{$c_f$} & correction factor for the distance\\
& from the integrating sphere\\
\noalign{\smallskip}\hline
\end{tabular}
\end{table}

For a first guess, the calibration factor $c_f$ does not depend on
wavelength.
When looking more closely one recognizes the differences in the angle
dependencies of the spectral response of the photodiode for different
wavelengths, e.g. at long wavelengths the effective cross section of
the diode becomes larger for flat angles, while at short wavelengths a larger
fraction of the light is being reflected at the surface for flat angles.
This means in our case that the calibration factor $c_f$ is
wavelength-dependent, since inside the sphere the diode sees light
coming from all angles, while in the dark box only steep angles are
arriving at the diode.
We tried to overcome this problem by measuring $c_f$ for wavelengths
in between 400~nm and 1000~nm, extrapolating for
wavelengths below 400~nm.
In this region the diode current is of the same order of magnitude as
the fluctuations of the dark current, so it is not possible to measure
the light intensity directly inside the dark box.

\begin{figure}
\begin{center}
\resizebox{\hsize}{!}{\includegraphics{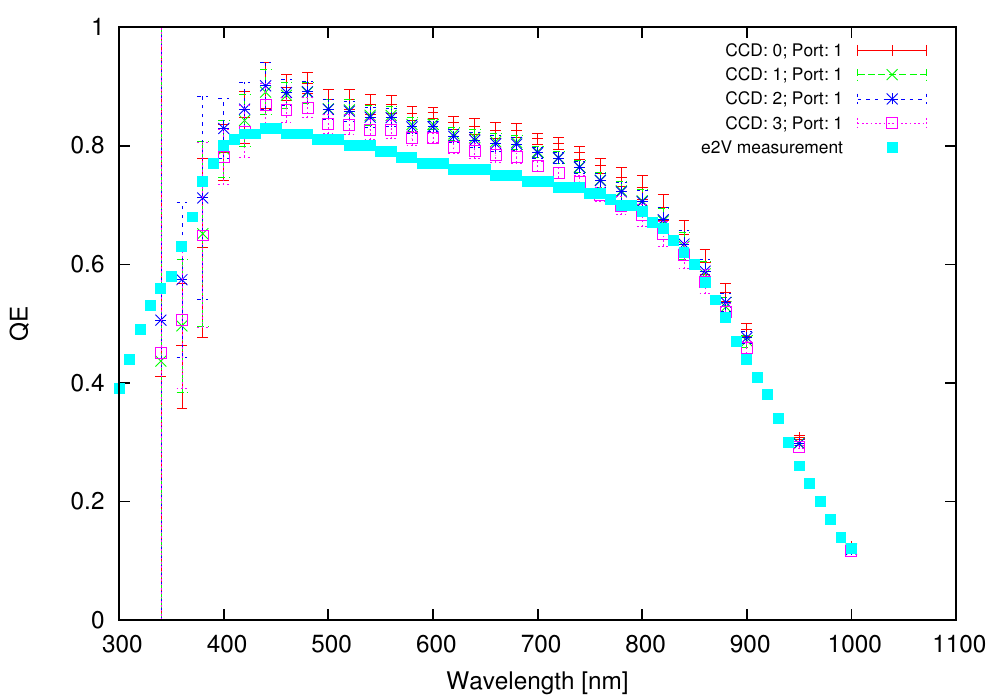}} 
\end{center}
\caption{Quantum efficiency of the four chips of the wide field imager
  measured in the lab of the USM (red, green, blue, purple), as well
  as the minimum guaranteed curve by e2v (cyan). \label{QE}}
\end{figure}

Figure~\ref{QE} shows the QE curves measured in the USM laboratory
(red, green, blue, purple, for the four chips) and by the manufacturer
(cyan).
It can clearly be seen that our lab measurement yields a slightly
higher QE than the one from e2v, nearly over the complete spectrum,
which makes sense as the curve from e2v is not an individual
detector measurement, but rather a minimum guaranteed curve.
The only exception is at wavelengths below 400~nm, where our
measurement yields lower results. These are, however, in agreement within
the error margins that are much larger in this region due to the very low 
photodiode currents.

\subsection{Filter transmission}

The transmissivity of the optical filters\footnote{The filters were 
manufactured by Omega Optical Inc, Brattleboro, USA} \citep[following the
SDSS-system: $ugriz$,][]{1996AJ....111.1748F} has also been measured in
our laboratory.
The measurement setup used the same light source and double
monochromator described in Sect.~\ref{sectQE}, but this time without
the integrating sphere.
Instead, the light from the monochromator is illuminating the
photodiode directly through the filter inside a dark box.
The diode current is measured once with filter and once without filter
to obtain the transmissivity.
This procedure was repeated for nine different equally distributed positions on the
filter, giving the average as the value for the filter transmission.
The measured transmission curves are shown in Fig.~\ref{Eff} (green
lines).

\subsection{Total efficiency}

In order to predict the on-sky performance of our camera, it is
necessary to determine the total efficiency of the system.
This includes:

\begin{itemize}
 \item Quantum efficiency of the detector (see Sect.~\ref{sectQE}).
 \item Transmission curve of each filter.
 \item Transmission of the field corrector, which consists of three
   lenses.
 \item Reflectivity of the primary, secondary and tertiary mirror.
 \item Extinction in the atmosphere, including the contributions from
   Rayleigh scattering, ozone absorption and aerosol scattering \citep{Bindel}.
\end{itemize}

With the total efficiency known one can calculate the number of
photons incident to the  Earth's atmosphere from the number of 
counts in a CCD image.

\begin{figure}
\begin{center}
\resizebox{\hsize}{!}{\includegraphics{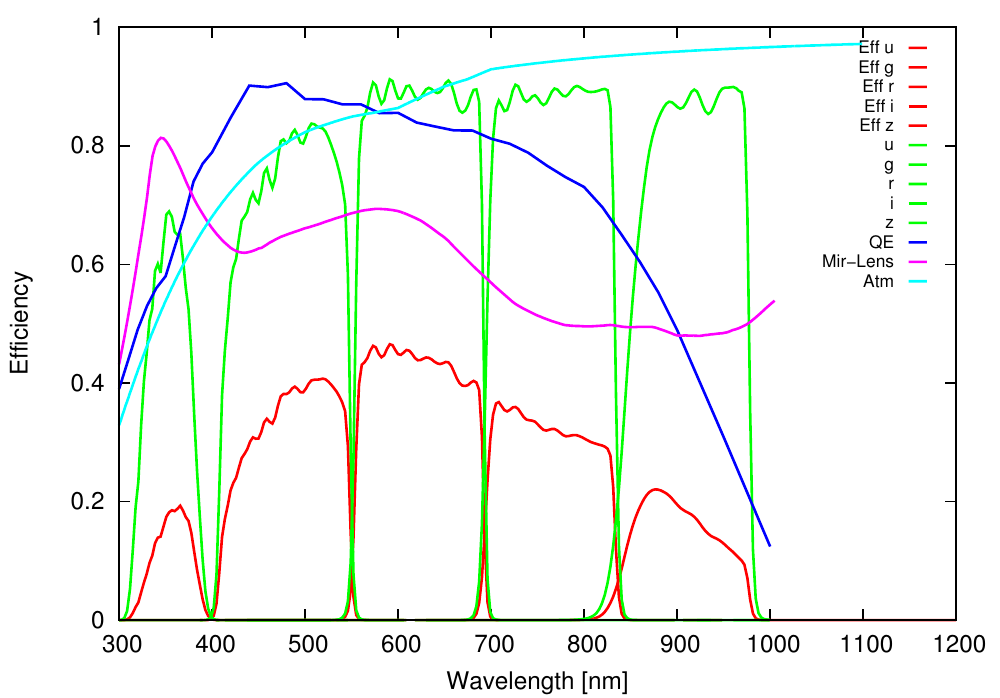}} 
\end{center}
\caption{Total Efficiency of the WWFI (CCD 0, Port 1) in \textit{ugriz} filters (red), filter
  transmission (green), QE of the detector (blue), combined corrector
  transmission and mirror reflectivity (all three, purple), combined
  atmospheric transmission at airmass unity (Rayleigh, ozone and aerosol, cyan).\label{Eff}}
\end{figure}

Since the statistical error of our QE measurement is very large at
wavelengths smaller than 400~nm (see Fig.~\ref{QE}) we decided to use
the manufacturer's QE below 400~nm and our own measurement above
this value as the ``true'' QE, as displayed in
Fig.~\ref{Eff} (blue curve).
Fig.~\ref{Eff} shows that the QE (blue curve) of the detector
is only of minor importance regarding the total efficiency, while major
contributions come from the mirrors (purple curve) at long wavelengths
and from the atmosphere (cyan curve) at shorter wavelengths. For the z-band, however,
the total efficiency is dominated by the steeply falling QE curve.

The contribution from ozone absorption is negligible (but has been
considered here), while Rayleigh and aerosol scattering both
contribute a significant fraction to the total efficiency, especially
at short wavelengths.
Since the aerosol abundance on Mt.\ Wendelstein is not known, we
followed \citet{Bindel} and assumed that the abundance is
comparable to that at Fred Lawrence Whipple Observatory on Mt.\
Hopkins at an altitude of 2617~m \citep{1975ApJ...197..593H}.

Table~\ref{SN} shows the limiting AB magnitudes\footnotemark (with
apertures of $1.1"$) of objects with which a 
signal-to-noise ratio of 5.0 can be achieved with a cumulative exposure
time of $1800~s$ (split into five exposures) taking into account all
system parameters and assuming unity airmass and a PSF with FWHM of $0.8"$.
\footnotetext{Following the definition by \citet{1983ApJ...266..713O}.}

\begin{table}[ht]
\caption{Predicted system throughput $Q$ and signal to noise ratio for
  a given AB magnitude in each filter for the WWFI for $5 \times
  360~s$ exposures, combined $\hat{=}1800~s$, PSF with FWHM $0.8"$, aperture $1.1"$ at airmass $1.0$.}
\label{SN}
\centering
\begin{tabular}{lccccc}
\hline\noalign{\smallskip}
waveband & $u$ & $g$ & $r$ & $i$ & $z$ \\
\noalign{\smallskip}\hline\noalign{\smallskip}
$Q$ & 0.201 & 0.363 & 0.415 & 0.325 & 0.155 \\

night sky AB & 22.80 & 21.90 & 20.85 & 20.15 & 19.26 \\

$S/N$ & 5.0 & 5.0 & 5.0 & 5.0 & 5.0\\

AB mag & 24.88 & 25.46 & 25.00 & 24.43 & 23.60\\

zero point & 24.25 & 25.41 & 25.36 & 24.87 & 23.54\\
\noalign{\smallskip}\hline
\end{tabular}
\end{table}

\subsection{Photometric zero points}

\label{sectZP}

The photometric zero point ($ZP$) is the magnitude of an object that produces
exactly one electron in a one second exposure in an instrument.
We measured the zero point of the WWFI with the first on-sky data
taken with the Wendelstein {\em Fraunhofer Telescope} of the globular cluster M13\footnote{Exposure times M13:
$u$: 60~s, $g$: 20~s, $r$: 10~s, $i$: 20~s, $z$: 40~s},
and with data from one night
of the Landolt standard star fields SA95\footnote{Exposure times SA95:
$u$: 60~s, $g$: 10~s, $r$: 10~s, $i$: 10~s, $z$: 20~s}, SA97\footnote{Exposure times SA97:
$u$: 30~s, $g$: 10~s, $r$: 10~s, $i$: 10~s, $z$: 10~s} and PG0918\footnote{Exposure times PG0918:
$u$: 60~s, $g$: 30~s, $r$: 30~s, $i$: 30~s, $z$: 30~s}.
In this section we describe the data analysis of the two
observations, present the results, compare them to each other and
compare them to theoretical predictions.

\subsubsection{Zero points from M13 data}

After bias subtraction and skylight flat-field calibration we used
SeXtractor \citep{1996A&AS..117..393B} on our images (in $u$, $g$,
$r$, $i$ and $z$ band) with a detection threshold of $3 \sigma$ for
4 contiguous pixels and
obtained aperture magnitudes with $1.5''$ diameter
(we chose this small aperture to avoid errors induced by crowding
effects, and extrapolated the magnitudes later on to an aperture of
$10.0''$ with 23 isolated bright stars in the outer region of the field).

We used the lists published by \cite{2008ApJS..179..326A} to identify
and match our stars, as well as for reference magnitudes to calculate
the zero point using the equation:

\begin{equation}
 ZP = m_\mathrm{lit} - m_\mathrm{inst} + AM \cdot \kappa - 2.5 \log(t_\mathrm{exp}) + 2.5 \log(g),
\label{eq_ZP}
\end{equation}
where $m_\mathrm{lit}$ is the magnitude from the catalog of
\citet{2008ApJS..179..326A} in the AB photometric system, $m_\mathrm{inst}$ is the (un-calibrated)
instrumental magnitude, $AM$ is the airmass which was $1.08$ in our
observation, $\kappa$ is the atmospheric extinction coefficient, for
which we took the average approximated values from \citet{Bindel}\footnotemark,
$t_\mathrm{exp}$ is the exposure time and $g$ is the gain of the detector (the estimated values
for the extinction are given in table \ref{ZP_tab}).
\footnotetext{Since we have only a single observation in each filter
  per airmass, we were not able to calculate the extinction.}

To minimize systematic errors we only accepted stars with:
\begin{itemize}
 \item Literature magnitude \textless 19.
 \item Distance from center of M13 \textgreater $350"$, in order to reject stars
   with bad photometry due to crowding effects in the center of the
   globular cluster.
 \item Magnitude error \textless 0.1 (from SeXtractor run). 
\end{itemize}

\begin{figure*}[ht]
\begin{center}
\includegraphics[width=0.32\textwidth]{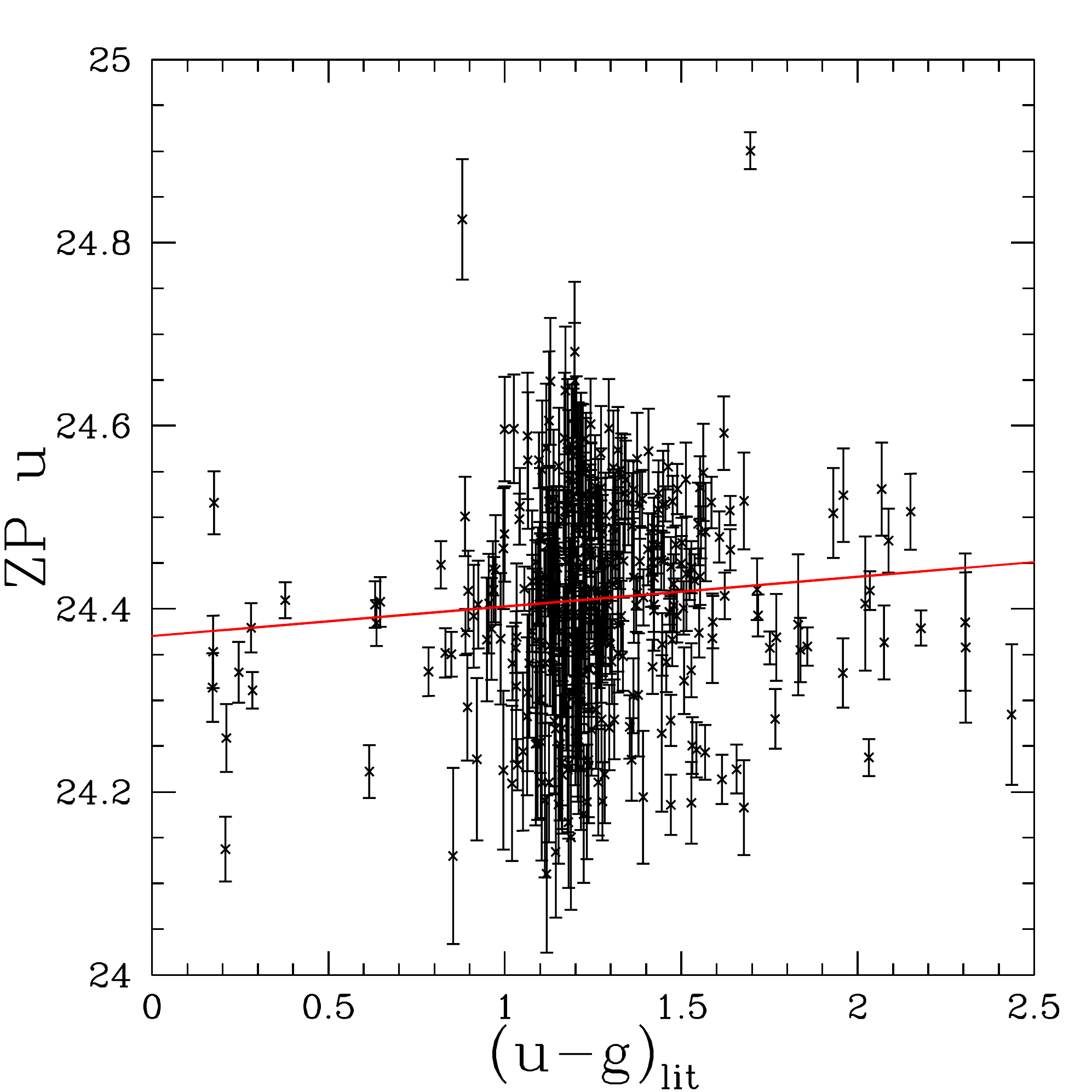} 
\includegraphics[width=0.32\textwidth]{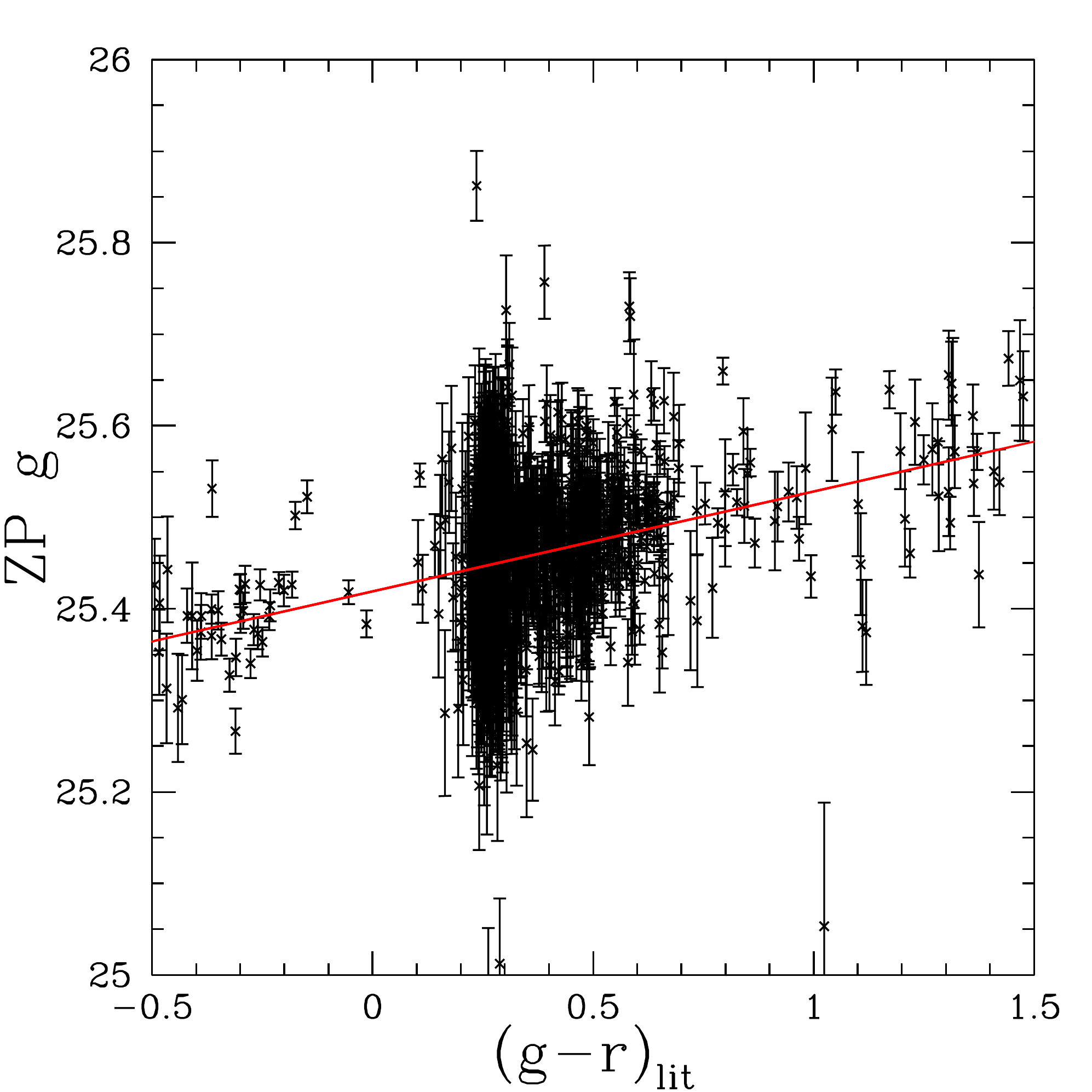} 
\includegraphics[width=0.32\textwidth]{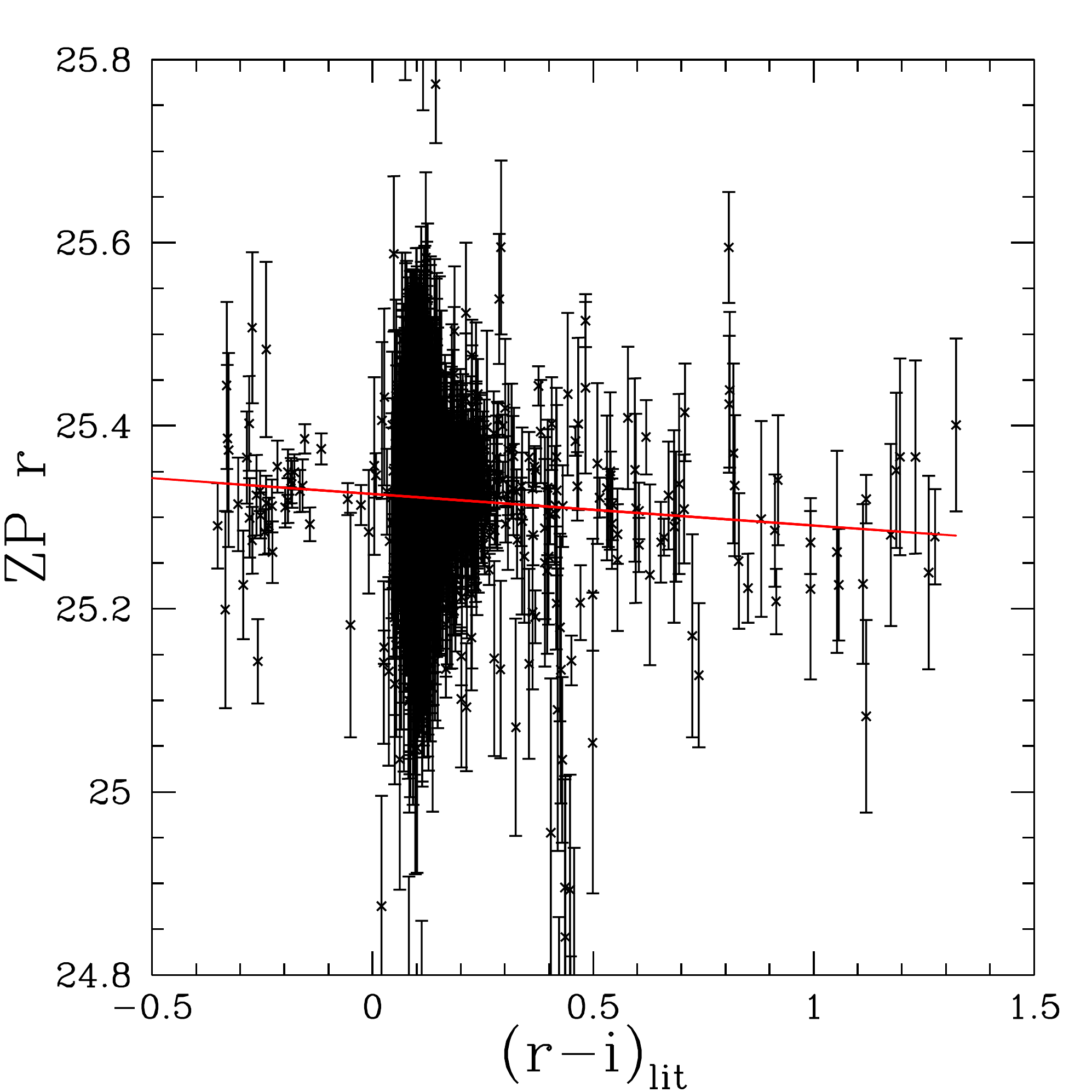} \\ 
\includegraphics[width=0.32\textwidth]{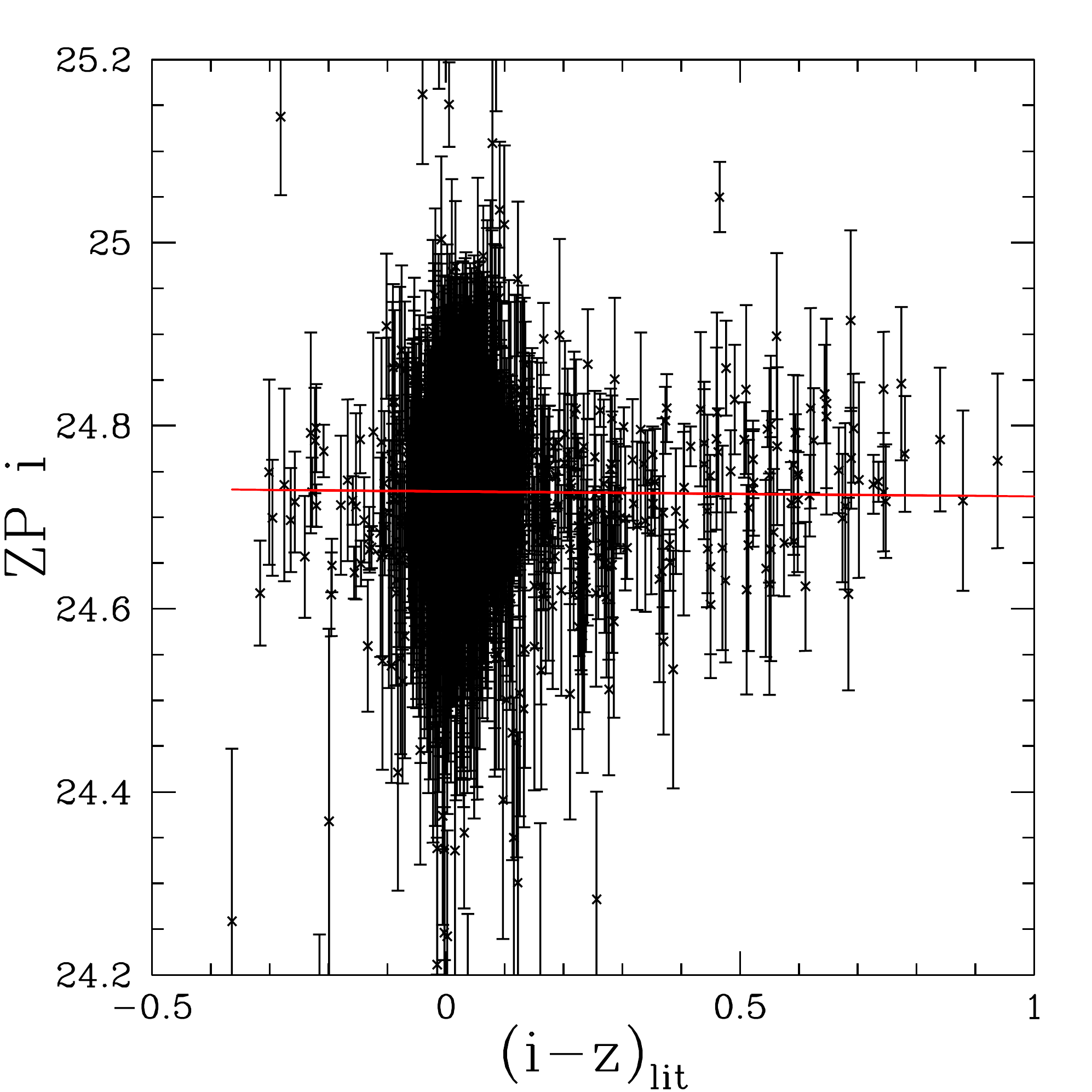} 
\includegraphics[width=0.32\textwidth]{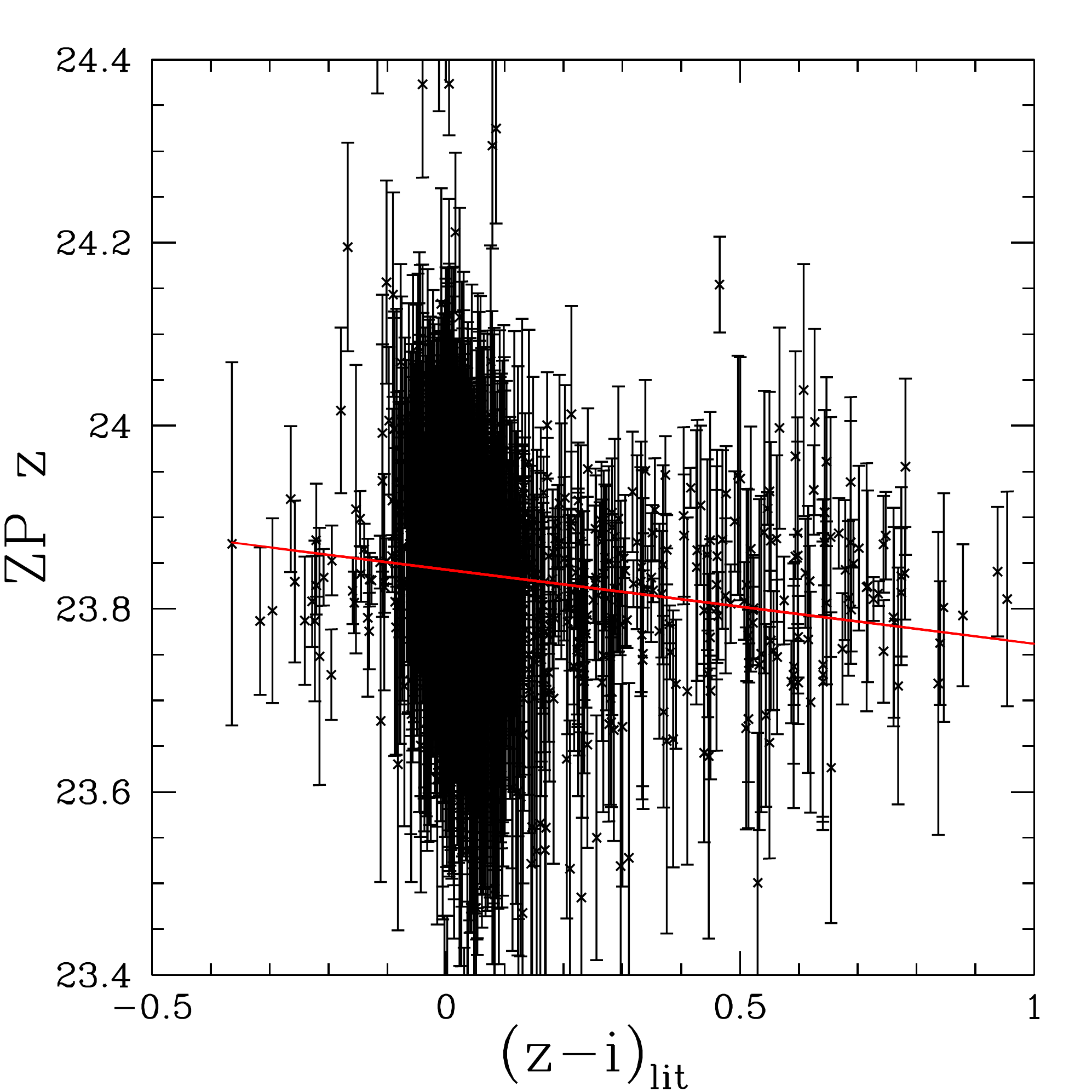} 
\end{center}
\caption{Zero point (average over all 4CCDs) from M13 data (in the AB-system) plotted vs.\
  (literature) color with linear fit to obtain the average zero point
  at color 0 and the corresponding color term.
  Top left: $u$ band $ZP$ vs.\ $u-g$, top right: $g$ band $ZP$ vs.\
  $g-r$, middle left: $r$ band $ZP$ vs.\ $r-i$, middle right: $i$ band
  $ZP$ vs.\ $i-z$, bottom: $z$ band $ZP$ vs.\ $z-i$; the scatter comes
  from the shallow depth of the observations and possibly also from variable sources
  in the catalog from \cite{2008ApJS..179..326A}.
\label{ZP_plot}}
\end{figure*}

To finally obtain the zero point in each filter we plotted the
individual zero point (each star) for each filter versus a
color (Fig.~\ref{ZP_plot}) and found that the resulting data can be
fitted linearly to obtain the color term as well:
\begin{equation}
 ZP(\mathrm{color})= a \cdot \mathrm{color} + ZP_0
\label{eq_colorterm}
\end{equation}
where $a$ is the color term and $ZP_0$ is the zero point at color 0.
The color term originates from comparing non-identical filters, but
since the magnitudes in the catalog from
\citet{2008ApJS..179..326A} are in the SDSS AB-system, we expect the
color terms to be very small (for identical systems, the color terms are
equal to 0).

Table~\ref{ZP_tab} shows the results of our zero point calculation.
We will discuss these at the end of Sect.~\ref{ZP_landolt}.

\subsubsection{Zero points from Landolt standard star fields data}
\label{ZP_landolt}

We used the Landolt standard star fields SA95, SA97 and PG0918
\citep{1973AJ.....78..959L,1983AJ.....88..439L,1992AJ....104..340L,2009AJ....137.4186L}
to measure the zero point again independently from the method explained above , with two
exposures per filter in SA95 and SA97 each and one exposure per filter in PG0918 for a total of five
airmasses for the calculation of the extinction coefficient.
The procedure of data reduction and application of photometry by
SeXtractor is the same as described in the previous section, with the
one exception that we used aperture diameters for photometry of
$10.0''$ from start, since we did not have to deal with a crowded
field here.
The main advantage over the previous method is the availability of
observations at multiple airmasses and thus the possibility to fit the
extinction coefficient for the particular night, rather than relying
on theoretical estimates for the atmospheric extinction.

The first step is to determine the extinction coefficient (in each
filter) by applying a linear fit to all stars that are detected at at
least two airmasses.
The extinction coefficient results from a global fit to all the star multiplets
simultaneously.
The slope of this linear fit (in magnitude over airmass) is equal to
the extinction coefficient.

We investigated the possibility of a variable extinction coefficient throughout the night
by comparing the magnitudes of stars dependent on time. We found a constant extinction coefficient
for each filter except the u-filter, where we estimated the systematic error from
varying extinction to be 0.05\ mag. We added this error to the flux error in our analysis in order to obtain a
better fit for the zero point in the u-filter.

\begin{figure*}[ht]
\centering
\includegraphics[width=0.48\textwidth]{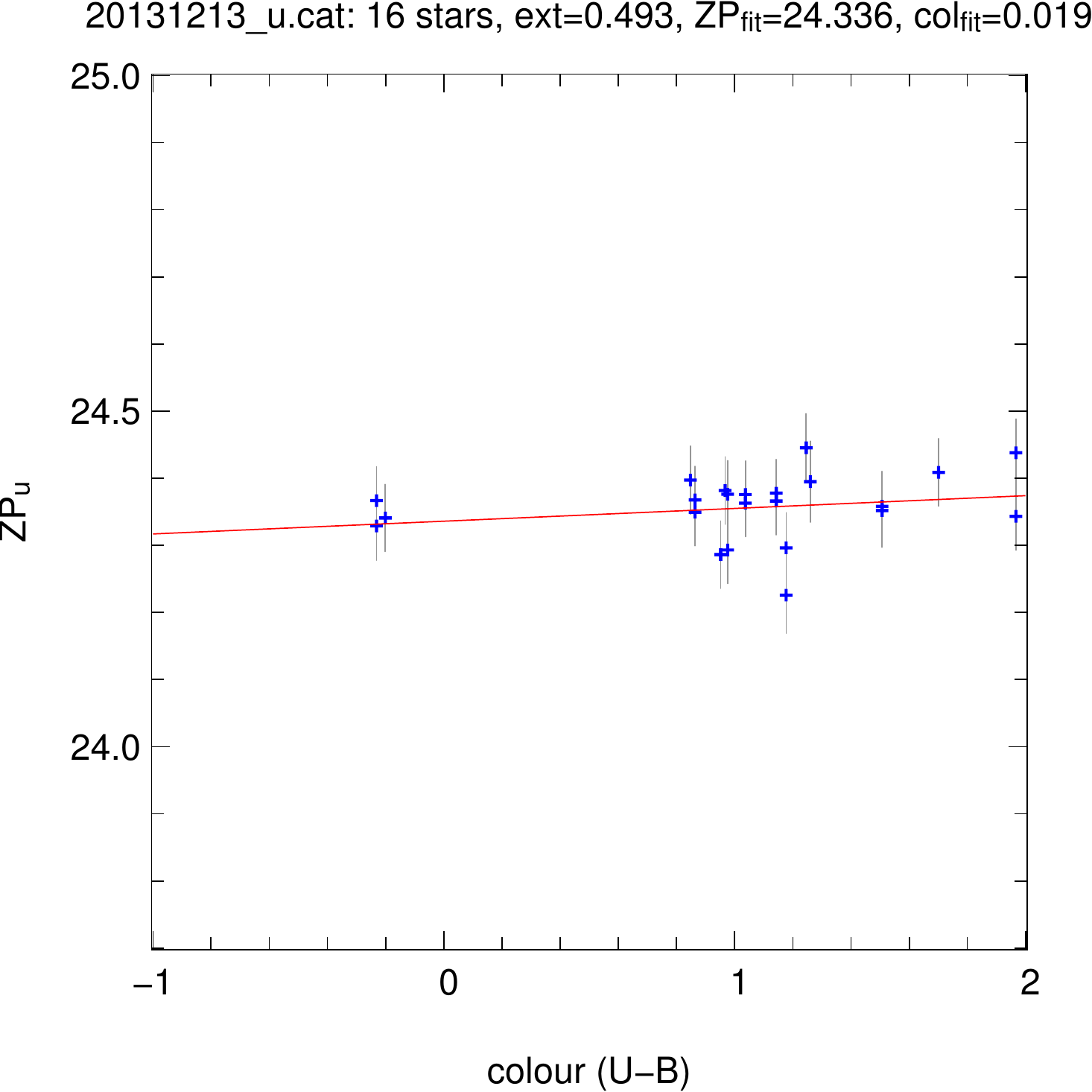}
\hspace*{2ex}
\includegraphics[width=0.48\textwidth]{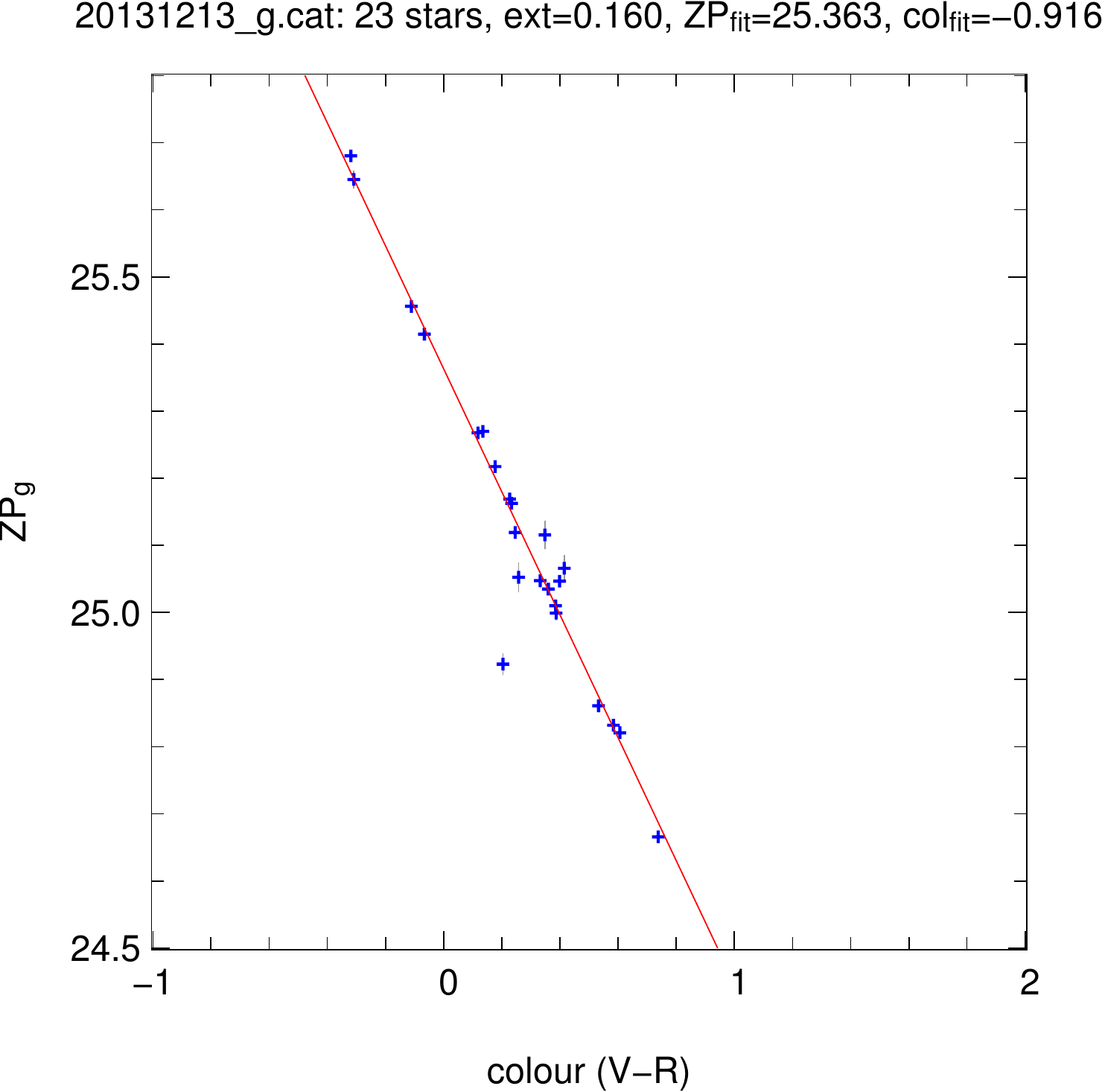}\\ 
\includegraphics[width=0.48\textwidth]{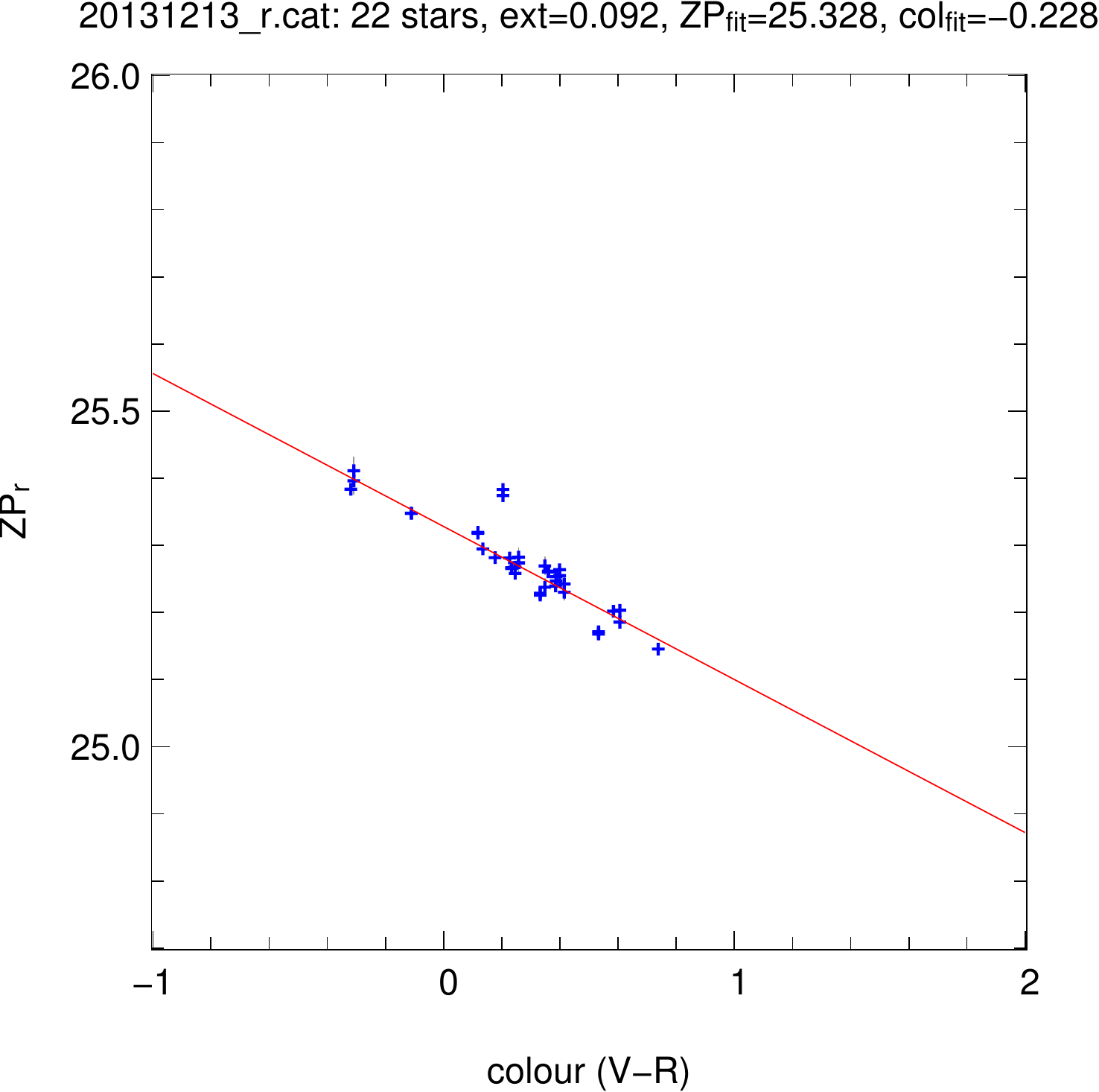} 
\hspace*{2ex}
\includegraphics[width=0.48\textwidth]{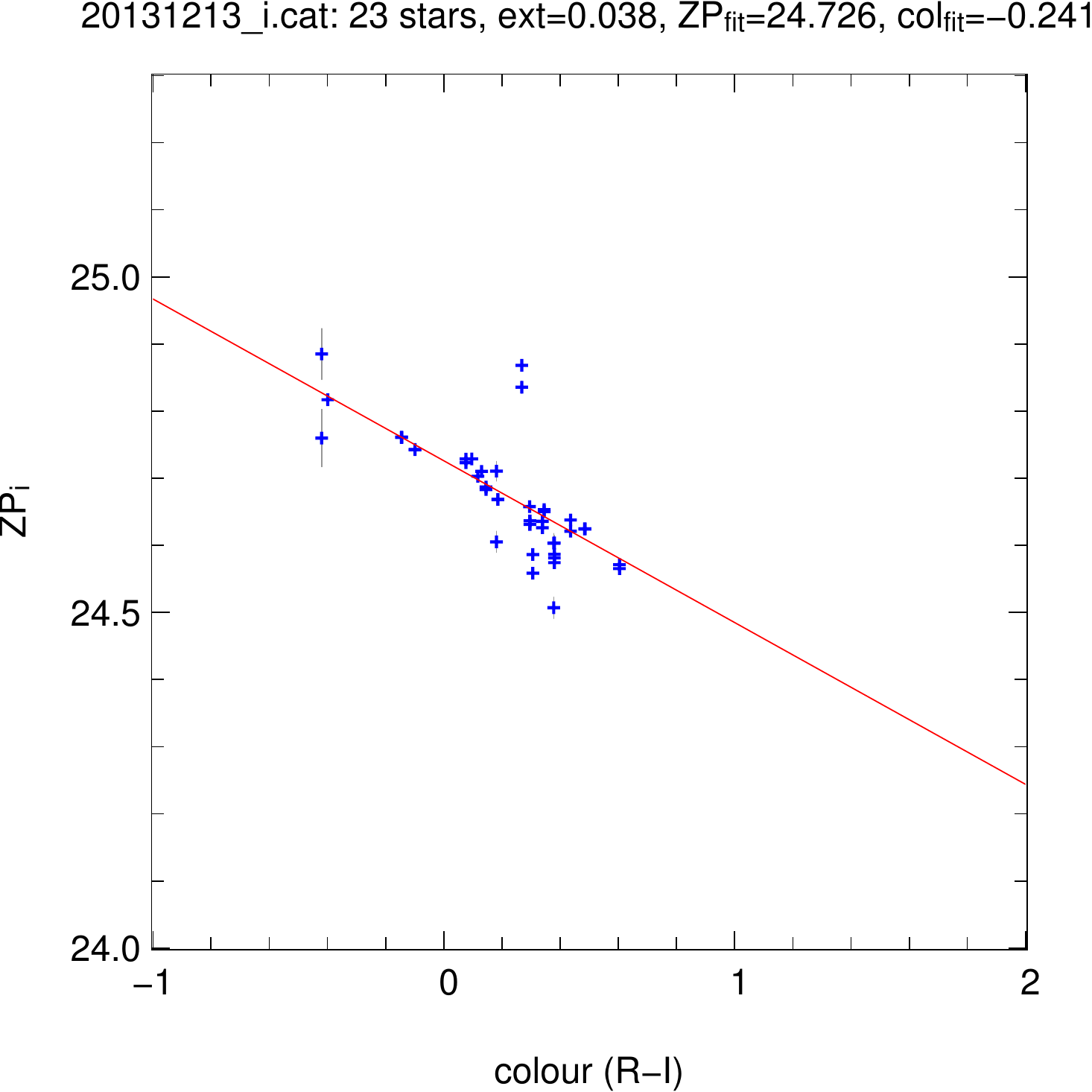} 
\caption{Zero points (in the AB-system) vs.\ colors from our standard
  star analysis.
  Top left: $u$ band $ZP$ vs.\ $U$-$B$, top right: $g$ band $ZP$ vs.\ $V$-$R$,
  bottom left: $r$ band $ZP$ vs.\ $V$-$R$ bottom right: $i$ band $ZP$ vs.\ 
  $R$-$I$.\label{ZP_plot_standards}}
\end{figure*}

After correcting for the extinction, our photometric catalogs are
matched with the standard star catalogs from
\citet{1973AJ.....78..959L,1983AJ.....88..439L,1992AJ....104..340L,2009AJ....137.4186L}.
Since the WWFI is using a filter set that is similar to SDSS
\citep[$ugriz$][]{1996AJ....111.1748F} and the Landolt catalog uses
Johnson-Morgan ($U$, $B$, $V$) \citep{1953ApJ...117..313J} and Cousins
($R_C,~I_C$) \citep{1976MmRAS..81...25C} filters, we have to
compare our magnitudes to the literature magnitudes taken from the
nearest (in terms of central wavelength) filter from the Landolt
catalog, which results in larger color terms.
Therefore, we compared our $u$ with $U$, our $g$ with $V$, our $r$ with $R$
and our $i$ with $I$.
We found that the filters are ``similar enough'' that a linear color
term is sufficient to correct for the differences
(Fig.~\ref{ZP_plot_standards}).
Unfortunately there is no adequate filter in the Johnson-Morgan and
Cousins system to compare our $z$ filter with, so we limited this
analysis to $u$, $g$, $r$ and $i$.
All magnitudes in the Landolt catalog, which are given in the
Vega-system, have been transformed to AB-magnitudes for our analysis.

In the near future, the photometry from the PanStarrs survey will be available for most of the northern
sky in the SDSS filter system, which will be a great opportunity to redo this kind of analysis without having the 
problem of converting between two photometric systems.

After the matching has been completed, we calculated a zero point for
each matched star via Eqn.~\ref{eq_ZP} and applied a linear fit to the
results in dependence of color (according to
Eqn.~\ref{eq_colorterm}), in order to determine the color term and
zero point at color 0.
Figure~\ref{ZP_plot_standards} shows the results of the linearly
fitted zero points over color, and Table~\ref{ZP_tab} summaries the
results of this measurement and the one from the previous section and
compares them to our theoretically predicted values based on our
laboratory results.
Table~\ref{ZP_tab} shows that there is an overall good agreement
between our two measurements, the deviations are always within the
margins of error.
The measured and observed values are in very good in agreement in the
$g$ and $r$ filter while in the $u$ filter the agreement is a little
worse, most probably due to the large uncertainties in the laboratory
calibration at short wavelengths arising from low illumination.
In the $i$ and $z$ filters the discrepancy is still a little larger
(0.14 and 0.15 respectively), and since the statistical error in in
this wavelength region is small, we conclude that this arises most
probably from systematic errors in the lab calibration.

\begin{table*}[ht]
\caption{Theoretical zero points as obtained by an exposure time
  calculator compared to the $ZP$s we measured on M13 data.
  All $ZP$s are in the AB photometric system.}
\label{ZP_tab}
\centering
\begin{tabular}{lccccc}
\hline\noalign{\smallskip}
waveband & $u$ & $g$ & $r$ & $i$ & $z$ \\
\noalign{\smallskip}\hline\noalign{\smallskip}
$ZP$ calculated & 24.25 & 25.41 & 25.36 & 24.87 & 23.96 \\
$ZP$ measured M13 & $24.37$ & $25.42$ & $25.33$ & $24.73$ & $23.84$ \\
$\Delta ZP$ M13 & $0.12$ & $0.072$ & $0.091$ & $0.091$ & $0.11$ \\
color term M13 & $0.032$  & $0.109$ & $-0.035$  & $-0.0055$  & $-0.081$ \\
color & $u$-$g$ & $g$-$r$ & $r$-$i$ & $i$-$z$ & $z$-$i$ \\
extinct. estimated M13 & 0.56 & 0.18 & 0.10 & 0.08 & 0.07\\
number of stars M13 & 382 & 1376 & 1482 & 1726 & 1807 \\
$ZP$ measured Landolt & $24.34$ & $25.36$ & $25.33$ & $24.73$ & \\
$\Delta ZP$ Landolt & $0.037$ & $0.018$ & $0.069$ & $0.031$ &\\
color term Landolt & 0.019  & -0.916  & -0.228  & -0.241 &  \\
color & $U$-$B$ & $V$-$R$ & $V$-$R$ & $R$-$I$ &  \\
extinction Landolt & 0.495 & 0.160 & 0.092 & 0.038 \\
\noalign{\smallskip}\hline
\end{tabular}
\end{table*}

\subsection{On-sky performance with stellar SEDs}

\begin{figure*}[ht]
\centering
\includegraphics[width=0.48\textwidth]{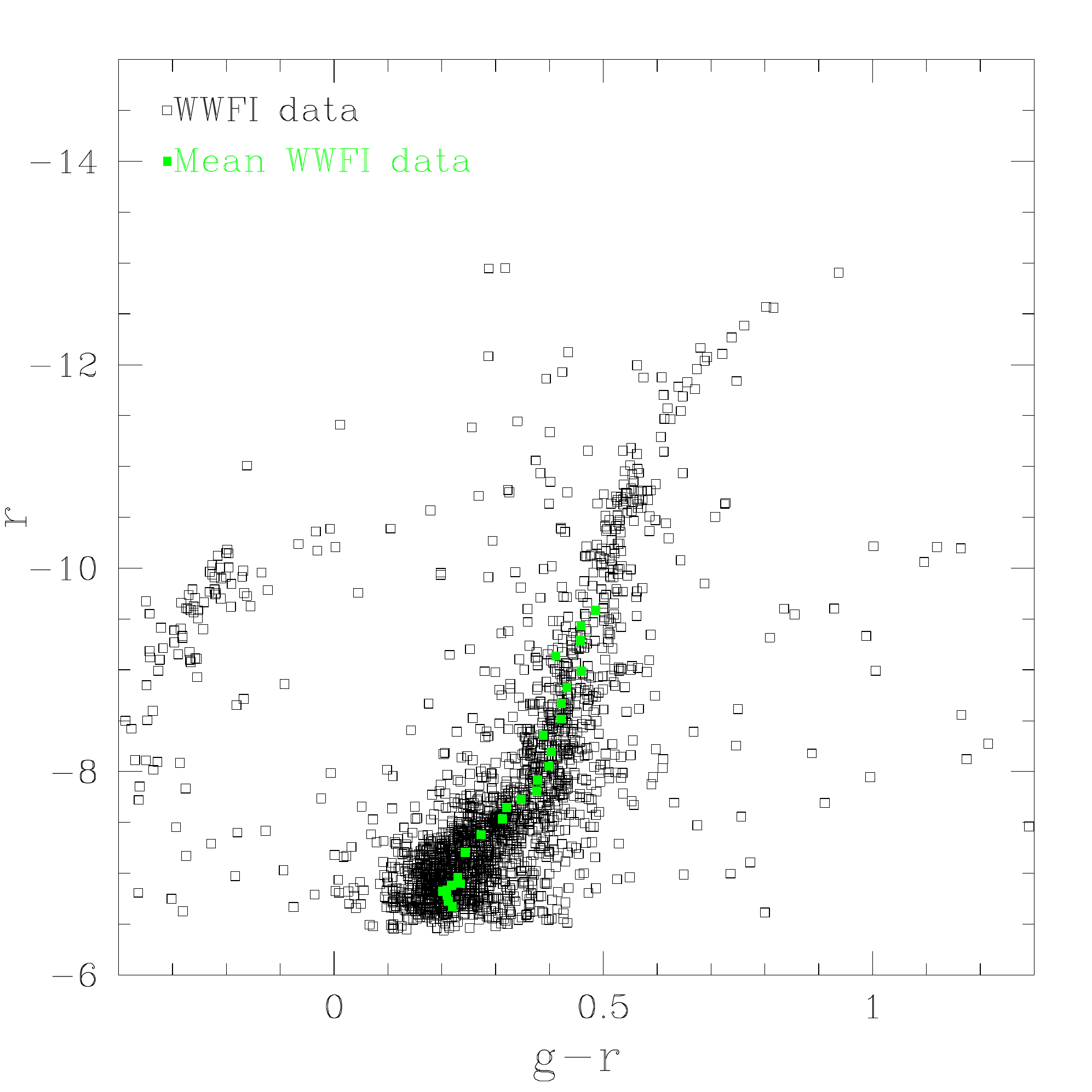}
\hspace*{2ex}
\includegraphics[width=0.48\textwidth]{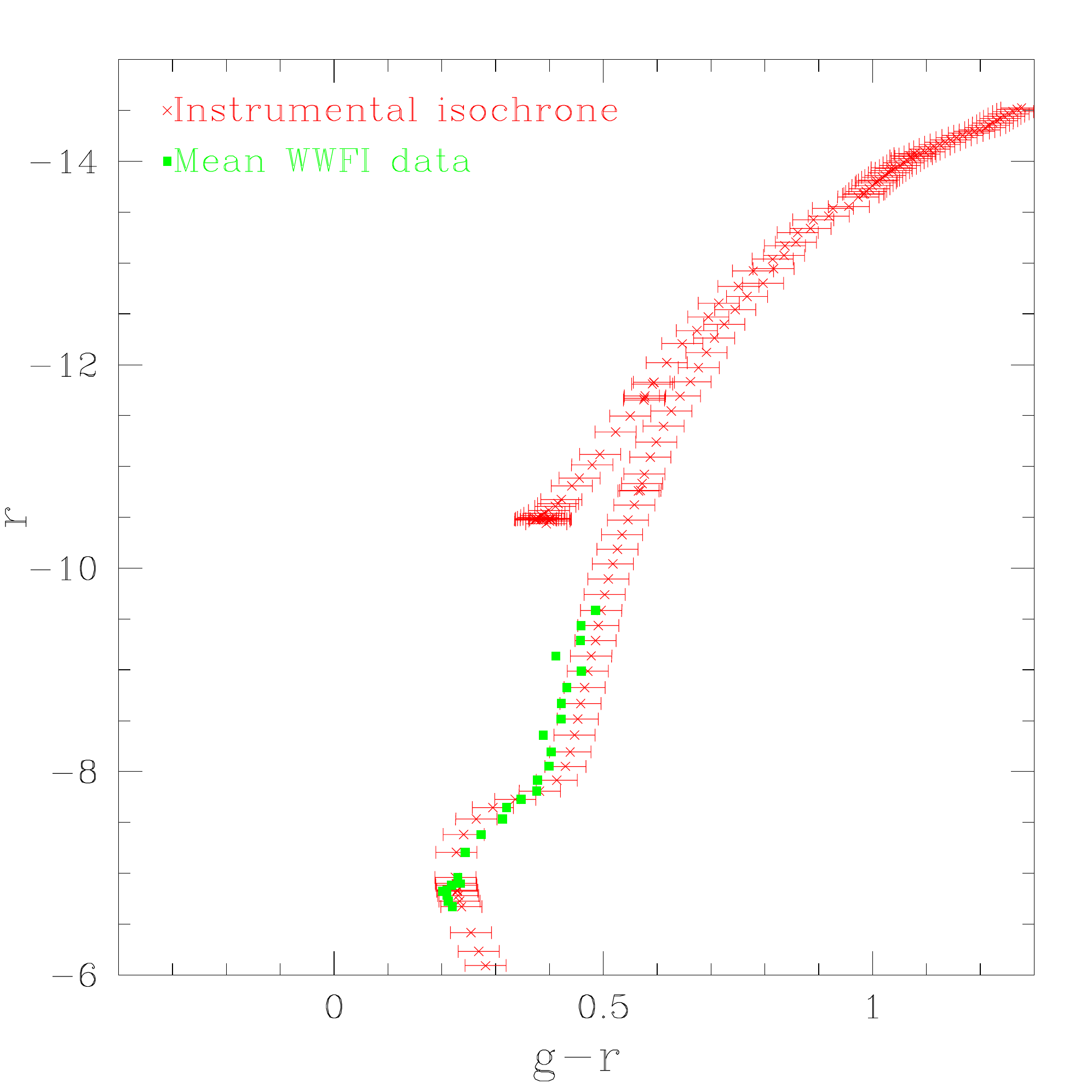} 
\caption{Instrumental color-magnitude diagrams, with $g-r$ color on
  the x-axis and $r$ band magnitude on the y-axis.
  Left: Black empty squares are data points from the observation and
  green filled squares represent the ridgeline (color-averaged) of
  these values.
  Right: Red crosses are expected magnitudes based on our lab-results (explanation see text)
  and green filled squares are again the ridgeline of the
  observational values.}
\label{ColMagInst}
\end{figure*}

The throughput of a system is defined as the amount of photons
detected by the instrument divided by the amount of photons incident
at the telescope aperture in a given filter.
Now we predict the instrumental magnitudes of objects in
our system depending on their spectral energy distributions (SEDs).
We use the same set of observations of M13 as described in
Sect.~\ref{sectZP}, since a globular cluster is very well suited for
this kind of analysis because it consists of stars of approximately the
same age and metalicity, thus of the same isochrone.
To obtain theoretical magnitudes for comparison we used the synthetic
stellar SEDs from Kurucz ATLAS 9 [as described in
\citet{1997A&A...318..841C}, available on the CD-ROM No.~13 of
\citet{1993KurCD..13.....K} based on the initial grid from
\citet{1979ApJS...40....1K}] and the isochrones from
\citet{2004A&A...422..205G}.
Since the Kurucz spectra are on a grid spaced by 0.5 in $\log(g)$
and by 250~K at low temperatures (and more coarsely at higher temperatures)
it is not possible to assign a separate SED to each entry of the
isochrone.
Thus, we interpolated linearly in $\log(g)$ and $\log(T_\mathrm{eff})$
to estimate the SED for each isochrone entry.
The so found SEDs were then convolved for each filter with the instrumental
efficiency curve measured in our lab (as presented in Fig.~\ref{Eff},
red curve) to find the instrumental
magnitudes these stars would have with our camera.
These magnitudes were then corrected for the distance modulus of M13
\citep[$14.44 \pm 0.06$ from][]{1992MNRAS.257..731B}
and for the interstellar extinction\footnotemark.
\footnotetext{From the \citet{2011ApJ...737..103S} recalibration of the
\citet{1998ApJ...500..525S} dustmap.}
These theoretical magnitudes are then plotted into a color-magnitude
diagram and compared to the observational data, as shown in
Fig.~\ref{ColMagInst}.
The black empty squares in the left panel represent the observational
data, the red crosses in the right panel are the expected magnitudes
(based on our lab data, the Kurucz-spectra and the isochrone) and the
green filled squares in both panels represent the ridgeline of the
observational values. We computed the ridgeline as the color-averaged values in
magnitude bins, each centered at the magnitude position of an (instrumental)
isochrone data point and a bin width equal to half the difference to
the neighboring isochrone data points.
At the bright end of the color-magnitude diagram of the globular
cluster the sequence is very sparsely populated.
In this region the objects scattered around the sequence (which are in
fact field stars not belonging to the globular cluster) would have a
large systematic impact on the averaging process and thus making it
very difficult to define a ridgeline.
Due to this reason we decided to apply a magnitude cut at the bright
end of the sequence (cut level depends on filter) and restricted this
analysis to the region where the sequence is densely populated.

\begin{table}
\caption{Color differences (RMS) for different color-magnitude
  combinations between the ridgeline of the measured values and the
  expected instrumental values.}
\label{ColMagDiff}
\centering
\begin{tabular}{lccccc}
\hline\noalign{\smallskip}
color and & $u-g$ & $g-r$ & $g-r$ & $g-i$ & $z-i$ \\ 
waveband & $u$ & $g$ & $r$ & $i$ & $z$ \\ 
\noalign{\smallskip}\hline\noalign{\smallskip}
difference [mag] & 0.083 & 0.037 & 0.030 & 0.063 & 0.057 \\
\noalign{\smallskip}\hline
\end{tabular}
\end{table}

Table~\ref{ColMagDiff} shows the root mean square of the color difference between
the ridgeline and the expected instrumental colors, for different
combinations of colors and magnitudes.
For all combinations we tested the differences are between 0.030 and
0.083. 

In this section we showed how well we can predict the
performance of our system using the calibration
measurements in the laboratory.
The numbers are compatible with the relative errors of our laboratory
calibration at the corresponding wavelengths, which shows that
there is no dominant systematic error.
The performance of this kind of prediction can be improved by using a
more sensitive lab calibration system (especially more sensitive at
short wavelengths).
Furthermore it would help to have observations at different airmasses
at hand, in order to be able to correct for atmospheric extinction
more accurately.

\subsection{Charge persistence}

Persistent charges, also called {\em residual images} can be divided
in two different forms:
Residual surface images (RSI) and residual bulk images (RBI).
RBI are only caused by photons with a high penetration depth, thus
they generally occur only when the chip is illuminated by radiation
with wavelengths greater than 700nm.
RSI can occur after illumination by any wavelengths.
RSI and RBI can be distinguished by their appearance:
RBI cause persistent charges only in the pixels that were illuminated,
while RSI cause the complete column (parallel to the readout direction) to
bleed.
If one observes bleeding columns with a spot somewhere which is
bleeding stronger than the rest of the columns, both RSI and RBI are
present.
A very detailed explanation of this effect can be read in \cite{janesick2001scientific}.

The detector of the WWFI is operated at a temperature of
$-115^{\circ}\mathrm{C}$ where the escape time of trapped charges should be large (\citet{1992ASPC...23....1J} state the decay time
to be exponentially dependent on chip temperature). Therefore, we have investigated
 whether the presence of residual images may hamper the performance of our detector.

\cite{janesick2001scientific}, \citet{1992ASPC...23....1J} and \citet{2012SPIE.8453E..1KB} state that one can get rid of residual (surface) images
in backside illuminated CCDs by inverting the clock voltage during readout, but unfortunately since we bought the detector system
as a ``black box'' we have no access to the detector electronics and are not able to adjust these parameters. So we have to live with that
problem and provide to the observer a useful workaround, which is what we try to do in this section.

\subsubsection{Method}

We used a mask with 64 small holes (hole diameter 1mm) in front of the
detector and a stabilized white LED to generate defined oversaturated
regions on the detector (16 per chip, 4 per port).
We oversaturated the spots
on the detector defined by the mask, then took a dark frame
immediately afterwards and repeated this procedure 10 times (of oversaturating and
taking a dark frame), where the only quantity that changes is the
exposure time of the dark frame.
In other words we were measuring the integrated value of decaying charges.
We also took as series of {\em real} dark frames (beforehand, without
residual images) for dark-subtraction.
The signal in each spot has been analyzed in a $20 \times 20$ pixel
box, while the diameters of the spots are approximately 100 pixels.
In order to characterize the effects of persistent charges, we
performed several measurements with varying parameters. These
parameters were:
\begin{itemize}
 \item the chip temperature,
 \item the amount of oversaturation that we defined as the charge in units of the full-well capacity, and
 \item the wavelength of the incident light.
\end{itemize}

\begin{figure*}[ht]
\centering
\includegraphics[angle=0,width=0.48\textwidth]{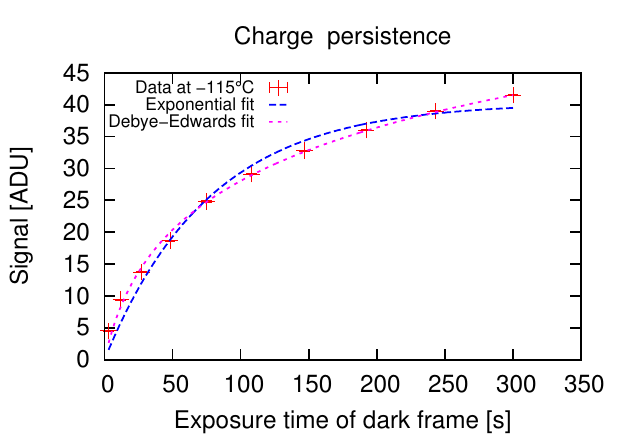}
\hspace*{2ex}
\includegraphics[angle=0,width=0.48\textwidth]{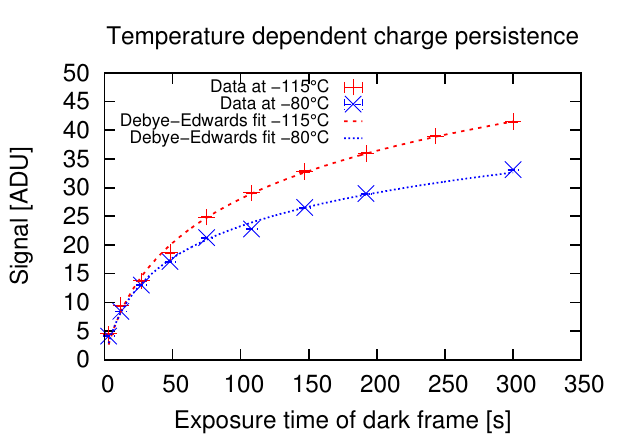}\\
\includegraphics[angle=0,width=0.48\textwidth]{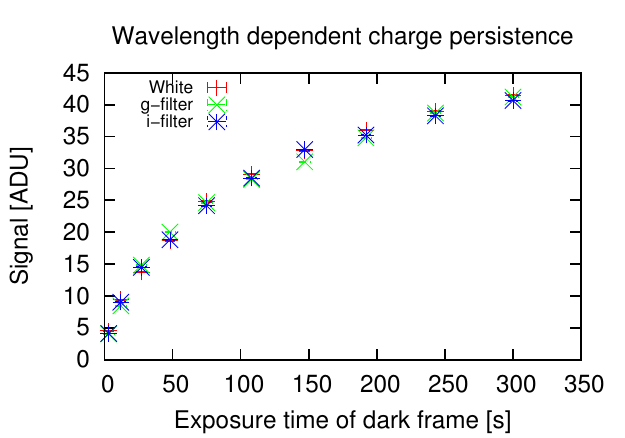}
\hspace*{2ex}
\includegraphics[angle=0,width=0.48\textwidth]{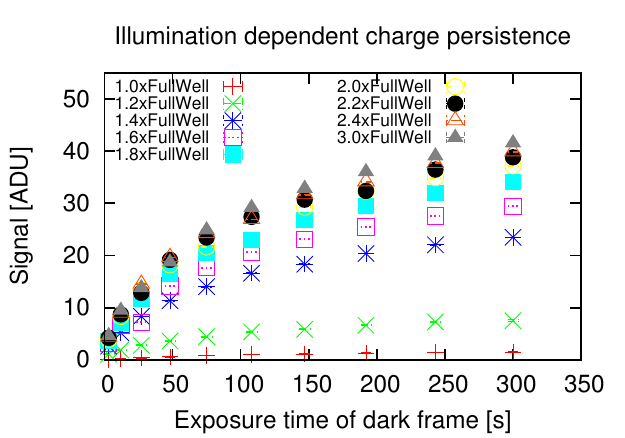}
\caption{Top left: Integral plot of persistent charges with
  exponential fit (blue) and Debye-Edwards fit (magenta).
  Top right: The same at $-115^{\circ}\mathrm{C}$ (red) and
  $-80^{\circ}\mathrm{C}$ (blue) with Debye-Edwards fits.
  Bottom left: The same for different wavelength regions: White light
  (red data points), SDSS $g$ filter (green) and $i$ filter (blue).
  Bottom right: The same for differing degrees of saturation, from $1
  \times$ full well capacity up to $3 \times$ full well. All plots with exception of
  the top right one are at $-115^{\circ}\mathrm{C}$.}
\label{ChargePers}
\end{figure*}

\subsubsection{Data analysis}
 
To quantify our results we plot the exposure time of the dark frame on
the horizontal axis and the total charge on the vertical axis, such that the
derivative of these functions represents the charge decay.
We tried to fit an integrated exponential function as well as an integrated Debye-Edwards
type decay function \citep[as proposed in][]{2012SPIE.8453E..1KB}
with a power-law exponent of 1, 
\begin{equation}
 F=\frac{A_0}{t+A_1}+A_2,
\end{equation}
to our data, where $F$ is the decaying charge, $A_0$ is the amplitude,
$A_1$ gives the variability with time and 
$A_2$ represents the contribution of the dark current to the signal.
The latter is equal to 0 in our case, since we subtracted a dark frame of the same 
exposure time from each image.

The top left plot of Fig.~\ref{ChargePers} shows the total charge in
the dark frame taken directly after saturation as a function of (dark)
exposure time.
CCD temperature was at $-115^{\circ}\mathrm{C}$ and the oversaturation
is three times the full well capacity.
The green curve shows an exponential fit and the blue curve shows a
Debye-Edwards fit.
Evidently, the fitting of the Debye-Edwards function works better,
which tells us that the decay of the charges happens not independently
for each electron, but is a function of the amount of trapped charges.
We assume that the electrostatic repulsion between the
trapped charges is the driving force of charge decay, but this
requires further investigation.
In the top right graph in Fig.~\ref{ChargePers} we show the persistent
charges for $-115^{\circ}\mathrm{C}$ (red) and $-80^{\circ}\mathrm{C}$
(green), with the result of faster decaying charges at higher temperatures, 
as expected. The bottom left plot of Fig.~\ref{ChargePers} shows the persistent
charges for different wavelength regions of incident light, i.e.\ 
white light, an SDSS $g$ filter and an SDSS $i$ filter\footnotemark
\footnotetext{$g$ filter: $\lambda = 4770~\AA, \Delta \lambda =
  1300~\AA$, $i$ filter: $\lambda = 7590~\AA, \Delta \lambda =
  1400~\AA$}.
The result is that the charge decay is independent from the wavelength
of the incident light, i.e.\ it does not matter how deep the radiation
penetrates into the pixel.
This proves that there are no residual bulk images, which should show
up only in the $i$-Filter, since only radiation with wavelengths greater than
$7000~\AA$ penetrates deep enough into the bulk to create them.
Our result is in agreement with \citet{janesick2001scientific}, who
states that residual bulk images do not show up in backside
illuminated devices.
In the bottom right graph of Fig.~\ref{ChargePers} the persistent
charges are plotted for several levels of oversaturation.
At light levels slightly above the full-well capacity the charge decay
time is strongly dependent on the light level, with a decreasing
dependence for higher illumination.

\begin{figure}
\centering
\resizebox{\hsize}{!}{\includegraphics[angle=0]{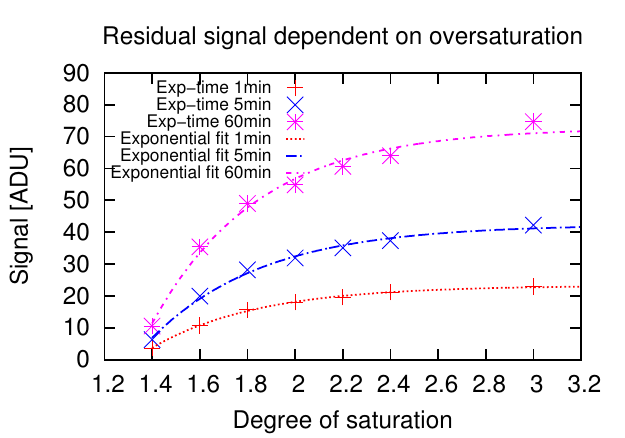}}
\caption{Residual signal vs.\ {\em oversaturation level} (defined as the charge per pixel in units
of the full-well capacity) for three
  different exposure times of the dark frame (red: 1 min, blue: 5
  min, magenta: 60 min), where the dark frame was taken immediately after
  saturation.}
\label{int_dependence}
\end{figure}

In order to characterize the dependence of the charge persistence on
the illumination level we plotted the residual signal vs.\ the {\em
  oversaturation level} (in units of full well capacity,
Fig.~\ref{int_dependence}).
The fact that these data are well fitted by an exponential function
means that there is a {\em worst case} (the asymptotic maximum), which
we can use for further treatments of the persistent charges.

The red, green, blue and magenta lines in the left plot of
Fig.~\ref{ChargePersDarkSky} show the charge persistence for different
{\em waiting} times between oversaturation and beginning of the
following exposure (with no wiping between the exposures).
Comparing these to the signal from the night sky background in the
current filter gives us the time we should mask out the oversaturated region.

\subsubsection{Dealing with persistent charges}

\begin{figure*}
\centering
\includegraphics[angle=0,width=0.48\textwidth]{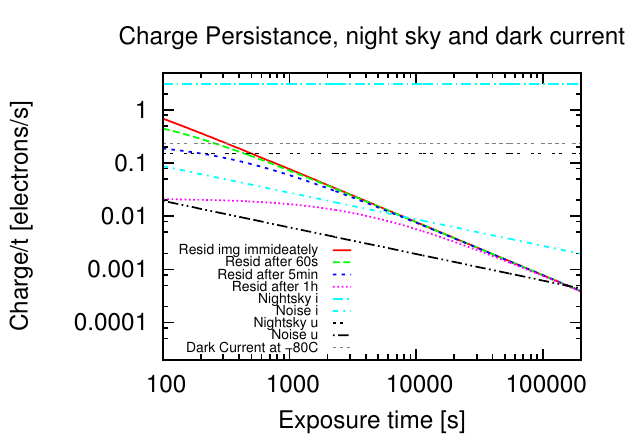} 
\hspace*{2ex}
\includegraphics[angle=0,width=0.48\textwidth]{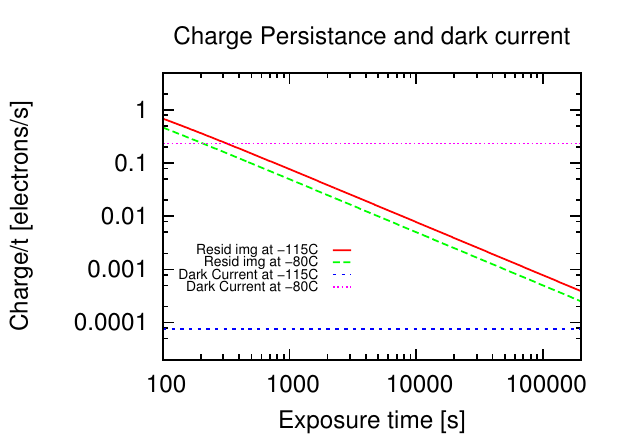}
\caption{Left: Plot of the persistent charge vs.\ exposure time of the
  subsequent image (not the integrated form as in fig \ref{ChargePers}, but in units of
  $\frac{e^-}{s}$), for an image
  taken immediately (red), 60 seconds (green), 5 minutes (blue) and 1
  hour (magenta) after saturation compared with the night sky background
  and its noise in the $i$ filter (cyan) and $u$ filter (black) and
  the dark current at a chip temperature of $-80^{\circ}\mathrm{C}$
  (the dark current at the operating temperature of
  $-115^{\circ}\mathrm{C}$ is not shown since it is extremely low at
  about $0.27 \frac{e^-}{h}$ and therefore not relevant).
  Both axes are logarithmic.
  Right: This plot has the same axes as the left plot, but shows
  persistent charges for two different temperatures (for the {\em
    immediate} case only) compared with the dark current at the same
  temperatures.}
\label{ChargePersDarkSky}
\end{figure*}

There are several possible ways of dealing with residual images:
\begin{enumerate}
 \item Run the detector at a higher temperature.
 \item Pre-flash (saturate) the detector before each sky exposure.
 \item Mask oversaturated regions for a defined amount of time.
 \item Prevent saturation, which is impossible for a wide field imager.
\end{enumerate}
The right graph of Fig.\ref{ChargePersDarkSky} clearly shows that
raising the temperature of the detector to accelerate the decay of
persistent charges is not an option  since the dark current rises by
a factor greater than $1000$ when changing the temperature from
$-115^{\circ}\mathrm{C}$ to $-80^{\circ}\mathrm{C}$.
Pre-flashing the detector would require a light source that
illuminates the detector area homogeneously.
Furthermore, pre-flashing is in principle the same as raising the dark
current and noise (by the needed amount to let residual images {\em disappear}
in the dark), so it is slightly preferable over a warmer detector, but
still not an ideal solution.

Masking of the oversaturated regions sounds like a method that is easy
to realize, but there are two issues one has to deal with:
First, one has to decide for how long one wants to mask the bleeding
regions.
By looking again at the left plot of Fig.~\ref{ChargePersDarkSky}, it
becomes clear that the amount of time has to depend on the filter of
the next exposure (since the level of the night sky background depends on the filter).
Second, it is not an easy task to decide which regions of the detector are saturated, since the
amount of analog to digital counts do not saturate by themselves,
but overflow and show lower values again at high illumination.
We decided to go for the masking solution, since it leaves most of the
detector area usable without adding an artificial signal (and noise).
Our task is now to find saturated regions:
Before the overflow effect sets in, the signal will be constantly
rising with illumination level, so if one finds a closed ring of pixel
maxima, one can tell for sure that everything inside this ring is
saturated.
We will use this method to find saturated regions, and flag these
regions in subsequent images (depending on the time interval between
the exposures and the filter used in the subsequent image).
The observer can then decide whether to discard the flagged regions.

\subsection{Charge transfer efficiency}
\label{sectCTE}

In this section we will present the results of our measurement of the
{\em charge transfer efficiency} (CTE) and characterize the dependence
of the CTE on the illumination level.
CTE is defined as the number of charges arriving at the target pixel
during a single shift, divided by the number of charges departing from
the original pixel.
Analogously, one defines the {\em charge transfer inefficiency} (CTI) as :
\begin{equation}
 \mathrm{CTI} = 1 - \mathrm{CTE}
\end{equation}

The effects that are responsible for the CTI are described for example in
\citet{janesick2001scientific}.
The CTI causes a distortion of image shapes along parallel and serial
readout direction (there is CTI in the serial register as well), since
the amplifier assigns the deferred charges to another pixel.
In fact, the deformation of images in both directions depends on the
parallel and serial CTE and on the amount of parallel and serial
shifts the charge undergoes until it reaches the readout amplifier.
An otherwise perfect PSF is no longer circular.
This may become important for applications where one wants to measure
image shapes, as in the analysis of weak gravitational lensing.
The effect of CTE on image shapes is further in investigated in
\citet{2010PASP..122..439R}.

Generally, CTE becomes better at higher illumination levels, since the
time constant of self-induced drift $\tau_\mathrm{SID}$ \citep{janesick2001scientific} 
becomes smaller for larger charge packets.
At very high signal levels (around half-well and higher), CTE can
again become worse because the time constant of fringing fields
$\tau_\mathrm{FF}$ becomes larger \citep{janesick2001scientific}.
Below that point, CTI can generally be described by a power law
dependent on signal level:
\begin{equation}
 \mathrm{CTI} = a \cdot \mathrm{signal}^b
\end{equation}
with $b$ generally $\sim -1.0 \dots -0.5$.

\subsubsection{Method}

There are several different methods for measuring the CTE.
A relatively straightforward method, which is both qualitatively and
quantitatively useful, is to take a series of flat field images at
different light levels and {\em overscan} the serial and parallel
registers to produce an image that is several pixels larger on both
axes than the actual detector.
If the CTE would be 1.0, one would measure just bias level in the
overscan region.
In real CCDs with CTEs slightly lower than 1.0 the light level in the
first row (or column, in case of serial register) of the overscan
region is slightly above the bias level depending on the value of the
CTE.
The CTE can be obtained as follows:
\begin{equation}
 \mathrm{CTE} = 1 - \frac{I_{n+1}}{I_n \cdot n}
\end{equation}
where $I_{n+1}$ is the mean intensity in the first row (column) of the
overscan, $I_n$ is the intensity in the last row (column) of the
active region and $n$ is the number of transfers necessary to read the
complete image (equal to the number of pixels per column (row)).
This method is called {\em Extended Pixel Edge Response}, and is
described in more detail in
\citet{mclean2008electronic} and \citet{janesick2001scientific} among several
other methods.

\subsubsection{Results}

\begin{figure}
\centering
\resizebox{\hsize}{!}{\includegraphics{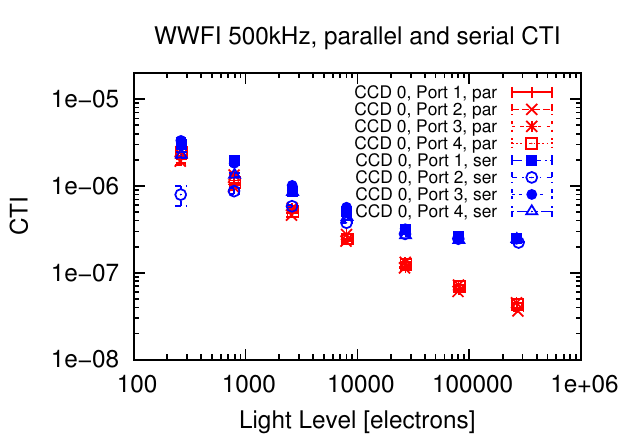}}
\vspace*{2ex}
\resizebox{\hsize}{!}{\includegraphics{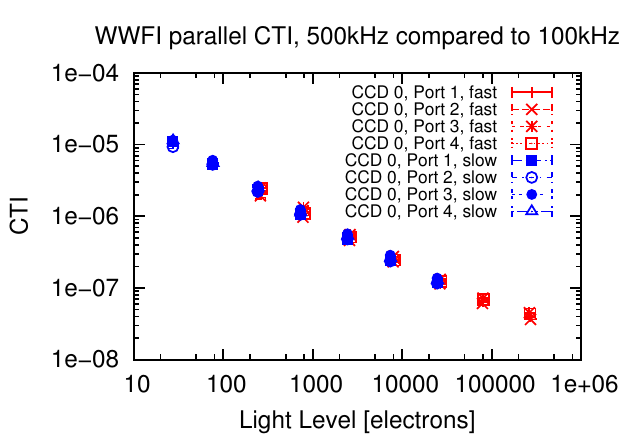}
\hspace*{20mm}
\includegraphics{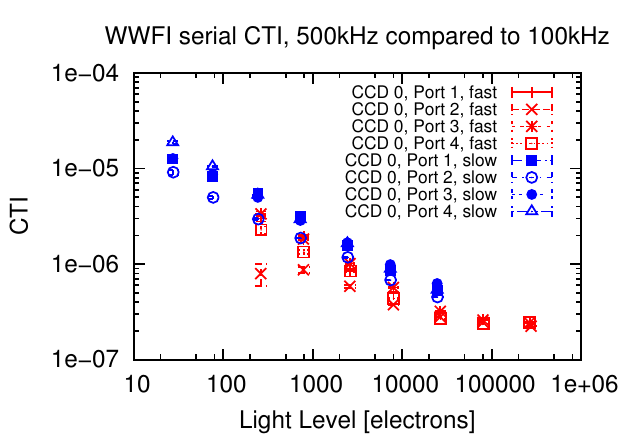}}
\caption{Top: Parallel CTI (red) compared to serial CTI (blue) in the
  500kHz readout mode in dependence of illumination, for one CCD of
  the WWFI.
  Bottom left: Parallel CTI in the 500kHz mode (red) compared to parallel
  CTI in the 100kHz mode. 
  Bottom right: Serial CTI in the 500kHz mode (red) compared to serial CTI in
  the 100kHz mode.
\label{CTE_fast_slow_par_ser}}
\end{figure}

The top graph in Fig.~\ref{CTE_fast_slow_par_ser} shows the parallel
CTI (red) compared to the serial CTI (blue) vs.\ the light level in
the last light sensitive line (column).
The data can be described by a power-law.
At illumination levels below $10000 e^-$ the serial CTI is higher than
the parallel one by a constant factor of approximately 1.5, at higher
illumination the serial CTI deviates from the power-law.
Usually that happens when the signal level approaches the full-well
capacity, but since we have not yet measured this quantity in the
serial register, we are not able to confirm this.
The middle graph in Fig.~\ref{CTE_fast_slow_par_ser} shows the
parallel CTI for the fast (red) and slow (blue) readout mode,
indicating that there is no difference.
The bottom graph in Fig.~\ref{CTE_fast_slow_par_ser} shows the
serial CTI for the fast (red) and slow (blue) readout mode. 
In this section we present only the results of one of the camera's CCDs
(number 0).
For the complete results and a comparison to the manufacturer's
results we refer the reader to App.~\ref{appA}.

We do not expect any problems with photometry as the CTI values of the
WWFI are very low.

\section{Comparison to similar systems}
\label{sectComp}

\begin{figure}
\centering
\resizebox{\hsize}{!}{\includegraphics{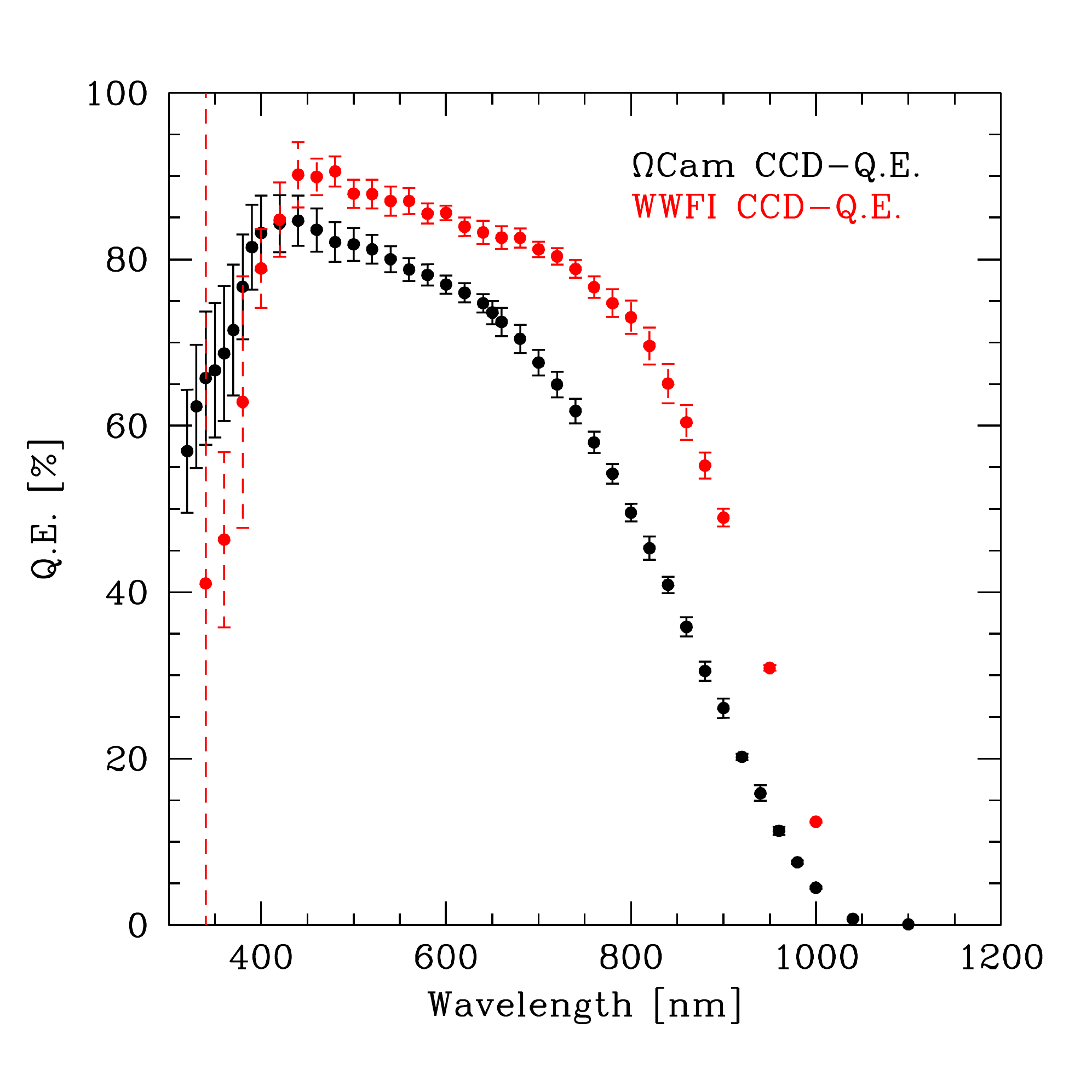}}
\caption{Quantum Efficiency of the OmegaCAM detector (black) compared to the WWFI (red).}
\label{QE_Omegacam_WWFI}
\end{figure}

\begin{table}
\caption{Comparison of WWFI with OmegaCAM and ESO-WFI. WWFI readout is via 4 ports per CCD, OmegaCam 
and ESO-WFI readout via 1 port per CCD.}
\label{WFI_comparison}
\centering
\begin{tabular}{lccc}
  \hline\noalign{\smallskip}
  instrument & \textbf{WWFI} & \textbf{OmegaCAM} & \textbf{ESO-WFI} \\
  \noalign{\smallskip}\hline\noalign{\smallskip}
   CCD type & e2v 231-84 & e2v CCD44-80 & e2v CCD44\\
  pixels & $\mathrm{8k \times 8k}$ &
  $\mathrm{16k \times 16k}$ & $\mathrm{8k \times 8k}$ \\
  field of view & $\mathrm{30' \times 30'}$ &
  $56' \times 56'$ & $34' \times 33'$ \\
  pixel scale & $0.2" / \mathrm{pixel}$ &
  $0.21" / \mathrm{pixel}$ & $0.24" / \mathrm{pixel}$ \\
  telescope & \raisebox{-1.2ex}[0pt][0pt]{2.0 m} &
  \raisebox{-1.2ex}[0pt][0pt]{2.6 m} & \raisebox{-1.2ex}[0pt][0pt]{2.2 m}\\
  aperture & & & \\
  gain & 5.81 or 0.69 & 0.54 & \\
  readout noise & $\mathrm{7.8 e^-}$ or $\mathrm{2.2 e^-}$ & 
  $\mathrm{5 e^-}$ &  $\mathrm{4.5 e^-}$  \\
  readout time & 8.5 s or 40 s & 29.5 s &  \\
  dark current & $0.27 e^- / \mathrm{h}$ & $0.54 e^- / \mathrm{h}$ & \\  
  \noalign{\smallskip}\hline
\end{tabular}
\end{table}

In this section we compare the parameters of the Wendelstein Wide
Field Imager with the ESO OmegaCAM \citep{2006SPIE.6276E...9I} at the VST survey telescope and
with the ESO-WFI \citep{1999Msngr..95...15B} at the 2.2~m Telescope at LaSilla.
Table~\ref{WFI_comparison} shows a comparison of the most important
parameters of the three wide field imagers.
In terms of pixel scale, all three imagers are compatible, the
OmegaCAM has a larger field of view since it has four times the amount
of pixels compared to ESO-WFI and to our camera.
One should point out that our imager has a significantly lower readout
noise when choosing the slow readout mode (and a compatible readout
time), while we could choose to have a much faster readout if we live
with a slightly higher readout noise.
The dark current of our camera at operating temperature is by a factor 2
lower than the dark current of the  OmegaCAM.

Figure~\ref{QE_Omegacam_WWFI} shows a comparison of the quantum
efficiency of the detector of OmegaCAM with the QE of the WWFI.
In the wavelength region above 450~nm the QE of the WWFI is higher by
approximately 5-10\%, while at short wavelengths the QE of the WWFI
seems to be lower but the QEs of the two detectors are in agreement
with each other in the margins of the errors of the WWFI
measurement in this region.

\section{Summary}
\label{conc}

We have presented the details about the mechanical construction of our
wide field imager as well as about the electromagnetic shielding and
the software.
Furthermore we have shown the details and results of our calibration
measurements in our laboratory as well as first on-sky data.
In Sect.~\ref{sectGain} we used the analytical method introduced in
\citet{2012SPIE.8446E..3PG} that successfully allows us to consider
data points at high count rates in our photon-transfer analysis even
when the photon noise of the masterflat begins to dominate.
We found reasonable results for the gain compared to the
manufacturer's estimation (Table~\ref{ReadoutNoise}).
Our quantum efficiency measurement in the laboratory shows only small
variations between the four CCDs and absolute values that are
slightly higher than the manufacturer's minimum guaranteed values (at
least at long wavelengths, while at short wavelengths our measurement errors 
are large).
We consider these values to be in good agreement.
We determined the photometric zero point of our system by two
different methods (an observation of a globular cluster with published
photometry and a standard star field) and found the results to be in
good agreement with each other (with exception of the $z$ filter where
we have only one result available), the dominant error source being
the atmospheric extinction which has been measured for the standard star analysis
but has been estimated for the globular cluster analysis.
The results are also generally in good agreement with theoretically
calculated values, with the exception of the $i$ and $z$ filters where
the dominant error source is assumed to be systematic errors in our
lab measurements.
We have shown that we can predict the on-sky performance of our
system with an accuracy between 0.030 and 0.083 in all colors.
To improve these numbers, a better lab equipment would be necessary,
especially a brighter calibration lamp.
We found out that the charge persistence in our detector is well
described by a Debye-Edwards law.
It varies with temperature and with illumination level, but is
independent from the wavelength of the incident light.
We were able to predict the amount of residual charge that remains on
the detector in dependence of time for the ``worst case'' of
oversaturation, which may be important for future observations.
We have shown that the CTE behaves as one would expect from low
values at low light levels to higher values at intermediate
illumination (it can be described by a power-law in this region quite
well), becoming slightly lower above half of full-well capacity.
In App.~\ref{appA} we show that results compare well to the values
determined by the manufacturer, with few exceptions for very few ports
only.
Finally in Sect.~\ref{sectComp} we found out that our system is very
well comparable to similar systems, ESO OmegaCAM \citep{2006SPIE.6276E...9I} 
and ESO-WFI \citep{1999Msngr..95...15B} in most
respects.
Our field of view is smaller than the FoV of OmegaCAM, but in terms
of QE and dark current our system is even better.

\begin{acknowledgements}
 
The authors thank Johannes Koppenhoefer and Mihael Kodric for their
support with the data reduction process with the WWFI.
Furthermore we thank Daniel Gruen for helpful discussions regarding
charge transfer efficiency.
We also thank Michael Schmidt and Christoph Ries, the night observers
at the Wendelstein Observatory for taking the necessary data for our
on-sky calibration. Michael Schmidt also took the responsibility for wiring
our imager, and we thank him for doing so.
We thank Wolfgang Mitsch for giving invaluable
advice on configuring the electronics of our camera.
We acknowledge the constructive discussion with Dietrich Baade and Olaf Iwert (ESO).
This research was supported by the DFG cluster of excellence “Origin
and Structure of the Universe” (www.universe-cluster.de).

\end{acknowledgements}

\bibliographystyle{aa} 

\appendix
\section{Charge transfer efficiency in more detail}
\label{appA}

\begin{figure*}
\begin{center}
\includegraphics[height=0.4\textheight]{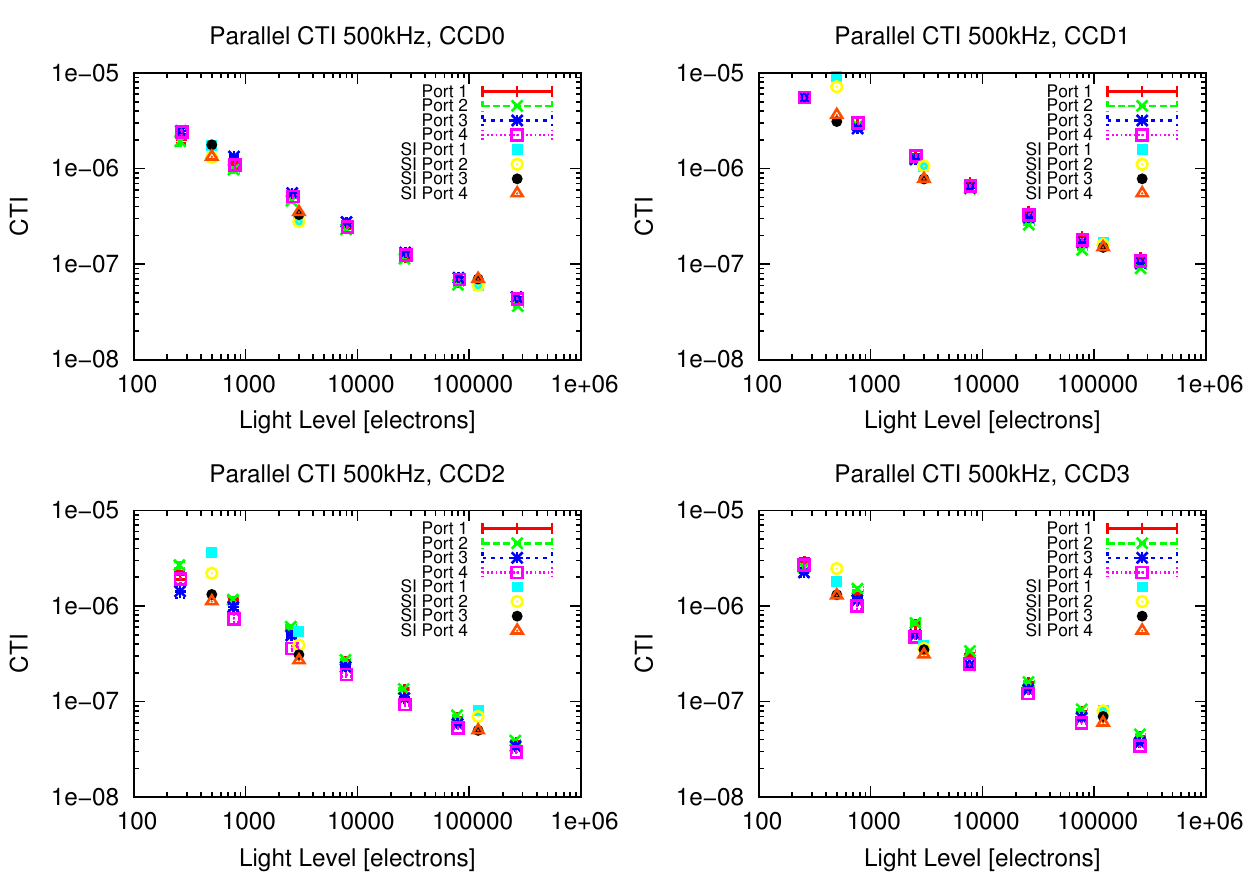} 
\end{center}
\caption{Parallel CTI for all four CCDs in the 500~kHz readout mode in
  dependence of illumination, compared to the values given by the
  manufacturer (SI).\label{CTE_fast_par}}
\end{figure*}

\begin{figure*}
 \begin{center}
 \includegraphics[height=0.4\textheight]{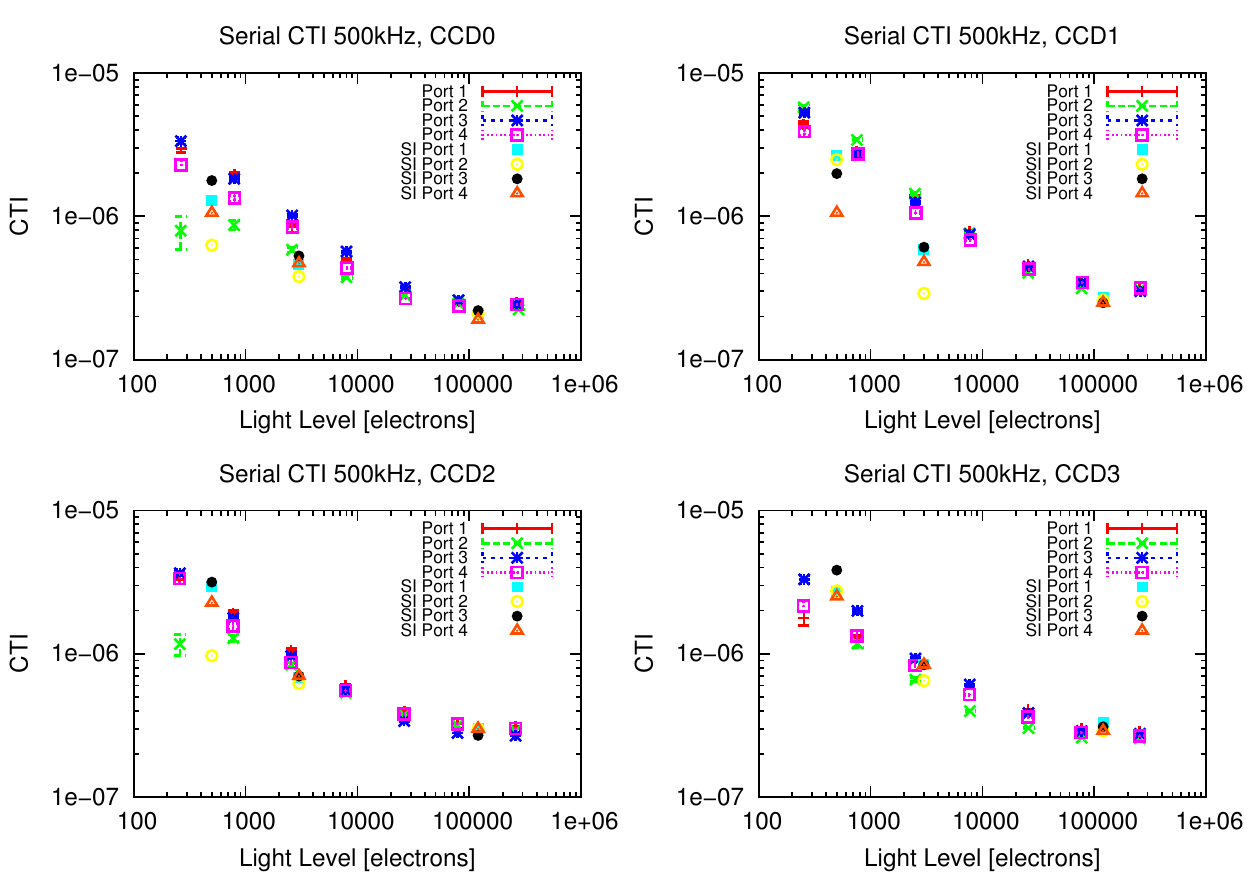}  
 \end{center}
 \caption{Serial CTI for all four CCDs in the 500~kHz readout mode in
   dependence of illumination, compared to the values given by the
   manufacturer (SI).\label{CTE_fast_ser}}
 \end{figure*}

In Sect.~\ref{sectCTE} we presented the results of the CTI measurement
in our laboratory, but we only showed results for one CCD (number 0).
In this appendix we will show the complete set of measurements for all
CCDs and compare them to the manufacturer's results.
Figure~\ref{CTE_fast_par} shows the parallel CTI for all four CCDs
compared to the values measured by Spectral Instruments.
(USM: red crosses, green, blue and magenta; SI: cyan, yellow, black
and red triangles).
The plots show overall good agreement between the two measurements
with few outliers in CCD1 and CCD2 (top right and bottom left) at low
signal levels, where the measurement performed by SI yields higher
values than our own results.
Figure~\ref{CTE_fast_ser} shows the same for serial CTI.
Here we can identify a few more outliers also at low signal levels,
but this time SI measures lower values than ourselves.
We trust our own measurements more than the SI measurements due to two
reasons:
First, our measurements are fitted by a power law, while the
measurements showing outliers are not, and second, the port-to-port
variations of our measurements (without outliers) are much smaller.

The reason why port 2 of CCDs 0 and 2 (green data points in top left
and top right of \ref{CTE_fast_ser}) show a lower CTI by approximately
factor of 3 at low signal levels are unknown to the authors.

\end{document}